\documentclass[11pt]{article}
\usepackage{jheppubmod}
\usepackage{color}
\usepackage[showonlyrefs]{mathtools}
\usepackage{psfrag}
\usepackage{array}
\usepackage{amssymb}
\usepackage{graphicx}
\usepackage{amsmath}
\usepackage{empheq}
\usepackage{slashed}
\usepackage{caption}
\usepackage[labelsep=quad]{subcaption}
\usepackage{epstopdf}
	
\usepackage{epsfig}
\usepackage[punctsep]{collref}
\newcommand{\be}{\begin{equation}}
\newcommand{\ee}{\end{equation}}
\collectsep[]{;}	
\usepackage{cite}
 
\allowdisplaybreaks

\def\DD{{\cal D}}

\def\II{{\cal I}}
\def\JJ{{\cal J}}
\def\KK{{\cal K}}
\def\LL{{\cal L}}
\def\MM{{\cal M}}
\def\NN{{\cal N}}
\def\OO{{\cal O}}

\def\SS{{\cal S}}

\def\VV{{\cal V}}
\def\WW{{\cal W}}

\def\qq{{\cal Q}}

\def\tts{{$tt^*$ }}

\def\Tr{{\rm {Tr}}}

\def\beq{\begin{equation}}
\def\eeq{\end{equation}}
\newcommand{\bea}{\begin{eqnarray}}
\newcommand{\eea}{\end{eqnarray}}

\title{
\tts equations, localization and exact chiral rings in 4d $\NN=2$ SCFTs
}

\author[a]{Marco Baggio,}
\author[b]{Vasilis Niarchos,}
\author[c,d]{Kyriakos Papadodimas}
\affiliation[a]{Institut fur Theoretische Physik, ETH Zurich,
CH-8093 Zurich, Switzerland.}

\affiliation[b]{Crete Center for Theoretical Physics
and Crete Center for Quantum Complexity and Nanotechnology,
Department of Physics, University of Crete, 71303, Greece.}
\affiliation[c]{Theory Group, Physics Department, CERN, CH-1211 Geneva 23, Switzerland.}
\affiliation[d]{on leave from the Centre for Theoretical Physics, University of Groningen,  The Netherlands.}
\emailAdd{baggiom@ethz.ch}
\emailAdd{niarchos@physics.uoc.gr}
\emailAdd{kyriakos.papadodimas@cern.ch}

\preprint{CCQCN-2014-38
\\ \hspace*{\fill}
CCTP-2014-15\\\hspace*{\fill}
CERN-PH-TH-2014-173}

\date{}

\abstract{We compute exact 2- and 3-point functions of chiral primaries in four-dimensional ${\cal N}=2$ 
superconformal field theories, including all perturbative and instanton contributions. 
We demonstrate that these correlation functions are nontrivial and satisfy exact differential equations with respect 
to the coupling constants. These equations are the analogue of the $tt^*$ equations in two dimensions. 
In the $SU(2)$ $\NN=2$ SYM theory coupled to 4 hypermultiplets they take the form
of a semi-infinite Toda chain. We provide the complete solution of this chain using input from
supersymmetric localization. To test our results we calculate the same correlation functions 
independently using Feynman diagrams up to 2-loops and we find perfect agreement  
up to the relevant order. As a spin-off, we perform a 2-loop check of the recent 
proposal of arXiv:1405.7271 that the logarithm of the sphere partition function in ${\cal N}=2$ SCFTs 
determines the K\"ahler potential of the Zamolodchikov metric on the conformal manifold.
We also present the \tts equations in general $SU(N)$ $\NN=2$ superconformal QCD theories 
and comment on their structure and implications.}


\begin{document}
\maketitle
\newpage

\section{Introduction}

In this paper we are interested in four-dimensional theories with $\NN=2$ superconformal invariance. 
There are many well known examples of $\NN=2$ quantum field theories (with or without a known 
Lagrangian description) that exhibit manifolds of superconformal fixed points (specific examples will be 
discussed in the main text). Although particular neighborhoods of these manifolds can sometimes be 
described by a conventional weakly coupled Lagrangian, the generic fixed point is a superconformal field 
theory (SCFT) at finite or strong coupling. It is of considerable interest to determine how the physical properties
of these theories vary as we change the continuous parameters (moduli) that parametrize these 
manifolds\footnote{The moduli of the conformal manifold in this paper should be distinguished from the moduli 
space of vacua, e.g. Coulomb or Higgs branch moduli, of a given conformal field theory.}. A well-studied 
maximally supersymmetric example with a (complex) one-dimensional conformal manifold
is $\NN=4$ super-Yang-Mills (SYM) theory. Large classes of examples are also known in 
theories with minimal ($\NN=1$) supersymmetry (see e.g. \cite{Leigh:1995ep}). Four-dimensional 
superconformal field theories with $\NN=2$ supersymmetry are particularly interesting because they are less 
trivial than the $\NN=4$ theories, but are considerably more tractable compared to the $\NN=1$ theories.

A particularly interesting subsector of $\NN=2$ dynamics is controlled by chiral primary operators. These
are special operators in short multiplets annihilated by all supercharges of one chirality. They form a chiral
ring structure under the operator product expansion (OPE). The exact dependence of this structure on 
the marginal coupling constants is currently a largely open interesting problem.

In two spacetime dimensions the application of the `topological anti-topological fusion' method gives rise 
to a set of differential equations, called \tts equations, which were employed successfully in the past 
\cite{Cecotti:1991me, Cecotti:1991vb} to determine the coupling constant dependence of correlation functions in
the $\NN=(2,2)$ chiral ring. An analogous set of \tts equations in four-dimensional $\NN=2$ theories was 
formulated using superconformal Ward identities in \cite{Papadodimas:2009eu}.\footnote{In a different 
direction, \tts geometry techniques have also been applied to higher dimensional quantum field theories more 
recently in \cite{Cecotti:2013mba, Vafa:2014lca}.} In four dimensions, however, it is less clear 
how to solve these differential equations without further input.

More recently, a different line of developments has led to the proposal that the exact quantum K\"ahler
potential on the $\NN=2$ superconformal manifold is given by the $S^4$ partition function of the theory 
\cite{Gerchkovitz:2014gta}. The latter can be determined non-perturbatively with the use of localization techniques 
\cite{Pestun:2007rz}. As a result, it is now possible to compute exactly the Zamolodchikov metric on the manifold of 
superconformal deformations of $\NN=2$ theories via second derivatives of the $S^4$ partition function. 
Equivalently, the two-point function of scaling dimension 2 chiral primaries is expressed in terms
of second derivatives of the $S^4$ partition function. We review the relevant statements in section 
\ref{ringgeneral}.

In the present work we take a further step and argue that, when combined with the \tts equations of 
\cite{Papadodimas:2009eu}, the exact Zamolodchikov metric is a very useful datum
that leads to exact information about more general properties of the chiral ring structure of 
$\NN=2$ SCFTs. Specifically, it provides useful input towards an exact solution of the \tts equations, which 
encodes the non-perturbative dependence of 2- and 3-point functions of chiral primary operators on the 
marginal couplings of the SCFT. In this solution, correlation functions of chiral primaries with scaling dimension 
greater than two are expressed in terms of more than two derivatives of the $S^4$ partition function. A review of the 
relevant concepts with the precise form of the \tts equations is presented in section \ref{ttreview}.

Such results can have wider implications. In subsection \ref{extremal} we demonstrate that a solution of the 
2- and 3-point functions in the $\NN=2$ chiral ring has immediate implications for a larger 
class of $n$-point `extremal' correlation functions. Moreover, it is not unreasonable to expect that 
2- and 3-point functions in the chiral ring may eventually provide useful input towards a more general solution 
of the theory using conformal bootstrap techniques.

In section \ref{sqcd} we demonstrate the power of these observations in an interesting well-known class of theories:
$\NN=2$ superconformal QCD defined as $\NN=2$ SYM theory with gauge group $SU(N)$
coupled to $2N$ fundamental hypermultiplets. This theory has a (complex) one-dimensional manifold
of exactly marginal deformations parametrized by the complexified gauge coupling constant 
$\tau = \frac{\theta}{2\pi} + \frac{4\pi i}{g_{YM}^2}$. For the $SU(2)$ theory, which has a single chiral ring 
generator, we demonstrate that the \tts equations take the form of a semi-infinite Toda chain\footnote{We remind 
that in certain two-dimensional examples with $\NN=(2,2)$ supersymmetry the \tts equations give a periodic Toda 
chain \cite{Cecotti:1991vb}.}. Solving this chain in terms of the $SU(2)$ $S^4$ partition function provides the exact 
2- and 3-point functions of the {\it entire} chiral ring. Unlike the ${\cal N}=4$ SYM case, where these correlation 
functions are known not to be renormalized \cite{Lee:1998bxa,D'Hoker:1998tz,D'Hoker:1999ea,Intriligator:
1998ig,Intriligator:1999ff,Eden:1999gh,Petkou:1999fv,Howe:1999hz,Heslop:2001gp,Baggio:2012rr}, in 
${\cal N}=2$ theories they turn out to have very nontrivial, and at the same time exactly computable, coupling 
constant dependence that we determine. In section \ref{sqcd} we also comment on the transformation properties of 
these results under $SL(2,{\mathbb Z})$ duality.
  
In the more general $SU(N)$ case, the presence of additional chiral ring generators makes the structure of the 
\tts equations considerably more complicated. A recursive use of the \tts equations is now less powerful and 
appears to require information beyond the Zamolodchikov metric (e.g. information about the exact 2-point 
functions of the additional chiral ring generators) which is not currently available. We present the $SU(N)$ \tts 
equations and provide preliminary observations about their structure.

Independent evidence for these statements is provided in section \ref{checks} with a series of computations
in perturbation theory up to two loops. Already at tree-level, agreement with the predicted results is a 
non-trivial exercise, where the generic correlation function comes from a straightforward, but typically 
involved, sum over all possible Wick contractions. We find evidence that there are compact expressions for 
general classes of tree-level correlation functions in the $SU(N)$ theory. The next-to-leading order 
contribution arises at two loops. We provide an explicit 2-loop check for the general correlation function 
in the $SU(2)$ $\NN=2$ superconformal QCD theory. As a by-product of this analysis we present 
a 2-loop check of a recently proposed relation \cite{Gerchkovitz:2014gta} between the quantum K\"ahler 
potential on the superconformal manifold and the $S^4$ partition function.

Some of the wider implications of the \tts equations and interesting open problems are discussed in section 
\ref{future}. Useful facts, conventions and more detailed proofs of several statements are collected 
for the benefit of the reader in four appendices at the end of the paper.
\vskip10pt
\noindent A companion note \cite{shortpaper} contains a consice presentation of some of the main results 
of this work with emphasis on the $SU(2)$ $\NN=2$ superconformal QCD theory.

\section{Marginal deformations and the chiral ring}
\label{ringgeneral}

\subsection{The chiral ring of ${\cal N}=2$ theories}
\label{chiralring}

The R-symmetry of 4d ${\cal N}=2$ SCFTs is $SU(2)_R\times U(1)_R$. We concentrate on (scalar) 
{\it chiral primary operators} defined as superconformal primary operators annihilated by all supercharges of
one chirality. These operators belong to short multiplets of type ``${\cal E}_{\frac{R}{2}(0,0)}$'' in the 
notation of \cite{Dolan:2002zh}\footnote{For an interesting recent discussion of other higher-spin chiral primary 
operators see \cite{Buican:2014qla}.}. As was shown there, these must be singlets of the $SU(2)_R$ and must have 
nonzero charge $R$ under $U(1)_R$. We work in conventions\footnote{In these conventions
the supercharges $\qq_\alpha^i$ have $U(1)_R$ charge equal to $-1$ and $\overline{\qq}_{\dot{\alpha}}^i$ 
have $+1$. The $\alpha,\dot{\alpha}$ are Lorentz spinor indices, while the $i$ is an $SU(2)_R$ index.} 
where the unitarity bound is
\begin{align}
\Delta \geq \frac{|R|}{2}~.
\end{align}
Superconformal primaries saturating the bound $\Delta = \frac{R}{2}$ are annihilated by all right-chiral 
supercharges $\overline{\qq}_{\dot{\alpha}}^i$. We call them chiral primaries and denote them by $\phi_I$. Their
conjugate, which obey $\Delta = -\frac{R}{2}$, are annihilated by $
\qq^i_\alpha$. We call them anti-chiral primaries 
and denote them  as $\overline{\phi}_I$. We write the 2-point functions of chiral primaries as
\be
\label{def2point}
\langle \phi_I (x) \overline{\phi}_J (0) \rangle = \frac{g_{I\overline{J}}}{|x|^{2\Delta}}~.
\ee
By the symbol $g^{\overline{J}I}$ we denote the inverse matrix i.e. $g_{I \overline{J}}g^{\overline{J}K} = \delta_I^K$.

It is well known that the OPE of chiral primaries is non-singular
\be
\label{defope}
 \phi_I(x)\, \phi_J(0) = C_{IJ}^K \, \phi_K(0) + \ldots~,
\ee
where $\phi_K$ is also chiral primary and $C_{IJ}^K$ are the chiral ring OPE coefficients \cite{Lerche:1989uy}.
We also define the 3-point function of chiral primaries 
\be
\label{def3point}
\langle \phi_I(x) \phi_J(y) \overline{\phi}_K(z) \rangle = \frac{C_{I J \overline{K}}}{|x-y|^{\Delta_I+\Delta_J-\Delta_K}
|x-z|^{\Delta_I+\Delta_K-\Delta_J}
|y-z|^{\Delta_J+\Delta_K-\Delta_I}}~,
\ee 
and we have the obvious relation between OPE and 3-point coefficients
\be
\label{ope3point}
C_{IJ\overline{K}} = C_{IJ}^L\, g_{L\overline{K}}
~.
\ee

So far we have defined the chiral ring for one particular ${\cal N}=2$ SCFT. In general, such SCFTs may have 
exactly marginal coupling constants. In that case the elements  of the chiral ring (i.e. the corresponding 2- 
and 3-point functions) will become functions of the coupling constants. The goal of our 
paper is to analyze this (typically non-trivial) coupling-constant dependence of the chiral ring.

\subsection{Marginal deformations}

We are interested in $\NN=2$ SCFTs with exactly marginal deformations. We parametrize the space of marginal 
deformations (conformal manifold), called ${\cal M}$ from now on, by complex 
coordinates $\lambda^i, \overline{\lambda}^i$. Under an infinitesimal marginal deformation the action changes by
\begin{equation}
\label{actiondeformation}
 S\rightarrow S + \frac{\delta \lambda^i}{4 \pi^2} \int d^4 x\, {\cal O}_i (x) 
 +  \frac{\delta \overline{\lambda}^i}{4 \pi^2} \int d^4 x\, \overline{{\cal O}}_i (x) 
 ~.
\end{equation}
It can be shown that the marginal deformation preserves ${\cal N}=2$ superconformal invariance, if and only if the 
marginal operators are descendants of (anti)-chiral primaries
with $\Delta = 2$ and $R=\pm 4$, more specifically
\begin{equation}
\label{marginalfromchiral}
{\cal O}_i = \qq^4\cdot \phi_i~, \qquad \overline{\cal O}_i = \overline{\qq}^4 \cdot \overline{\phi}_i~,
\end{equation}
where $\phi_i$ is chiral primary of charge $R=4$. The notation 
${\cal O}_i = \qq^4 \cdot\phi_i$ means that ${\cal O}_i$ can be written as the nested (anti)-commutator
of the four supercharges of left chirality. Their Lorentz and $SU(2)_R$ indices of the supercharges are 
combined to give a Lorentz and $SU(2)_R$ singlet. The overall normalization of factors of 2 etc. is
fixed so that equation \eqref{f192} holds. Notice that since the $\qq$'s have $U(1)_R$ charge equal to 
$-1$ the marginal operators are $U(1)_R$ neutral, as they should.

From now on in this section and the next
we use lowercase indices $i,j,...$ to indicate chiral primaries of R-charge equal to  $\pm4$. These are special since, via 
\eqref{marginalfromchiral}, they correspond to marginal deformations. We use uppercase indices $I,J,..$ to denote 
general chiral primaries of any $R$-charge.

The Zamolodchikov metric is defined by the 2-point function\footnote{Notice that 2-point functions of the form $\langle {\cal O}_i {\cal O}_j\rangle$ or
$\langle \overline{\cal O}_i \overline{\cal O}_j\rangle$ are zero, as can be easily shown by superconformal Ward identities.}
\be
\label{zdef}
\langle {\cal O}_i (x) \overline{{\cal O}}_j(0)\rangle = \frac{G_{i\overline{j}}}{|x|^8}
~.
\ee
The conformal manifold ${\cal M}$ equipped with this metric is a complex K\"ahler manifold 
(possibly with singularities).
The corresponding ``metric'' for the chiral primaries is
\be
\label{ctwopoint}
\langle \phi_i (x) \overline{\phi}_j(0)\rangle = \frac{g_{i\overline{j}}}{|x|^4}~.
\ee
We define the normalization of \eqref{marginalfromchiral} in such a way that 
$\langle {\cal O}_i (x) \overline{{\cal O}}_j(0)\rangle 
= \nabla^2_x \nabla^2_x \langle \phi_i (x) \overline{\phi}_j(0)\rangle$,
which implies 
\be
\label{f192}
g_{i\overline{j}} = \frac{G_{i\overline{j}}}{192}
~.
\ee

\subsection{The exact Zamolodchikov metric from supersymmetric localization}

In \cite{Gerchkovitz:2014gta} it was shown that the partition function of an $\mathcal{N} = 2$ theory on the 
four-sphere $S^4$, regulated in a scheme that preserves the massive supersymmetry algebra $OSp(2|4)$, 
computes the K\"ahler potential for the Zamolodchikov metric. The result is
\be
\label{metrick}
G_{i\overline{j}} = \partial_i \partial_{\bar j} {\cal K}
~,
\ee
where\footnote{In \cite{Gerchkovitz:2014gta} the marginal operators are normalized in a different way, namely 
$\mathcal{O}_{here} = 4 \mathcal{O}_{there}$, so various coefficients have been adjusted accordingly. For 
instance this explains the factor $192=12\times 4 \times 4$ as opposed to $12$ in \cite{Gerchkovitz:2014gta}.}
\be
\label{KZP}
{\cal K} = 192\log Z_{S^4}
~.
\ee
Combining this result with \eqref{f192} we conclude that
\be
\label{g2fromZ}
	g_{i\overline{j}} = \partial_i \partial_{\bar j} \log Z_{S^4}
~.
\ee
The partition function $Z_{S^4}$ can be computed exactly for a certain class of ${\cal N}=2$ SCFTs, using 
supersymmetric localization \cite{Pestun:2007rz}. Via \eqref{g2fromZ} this immediately provides the 2-point 
functions of chiral primaries with scaling dimension $\Delta = 2$. 

Our strategy will be to use these 2-point functions and the \tts equations that we derive in the following section to 
compute the 2-point functions of chiral primaries of higher R-charge. In turn, this will allow us to compute the 
{\it exact, non-perturbative} 3-point functions of chiral primaries over the conformal manifold.

\section{\tts equations in four-dimensional $\NN=2$ SCFTs} 
\label{ttreview}

In this section we review the analogue of the \tts equations for 4d $\NN=2$ SCFTs, which were derived in 
\cite{Papadodimas:2009eu}. We omit proofs, which can be found there.

\subsection{\tts equations and the connection on the bundles of chiral primaries}
\label{ttmaineqs}

We parametrize the conformal manifold ${\cal M}$ by complex coordinates $\lambda^i, \overline{\lambda}^i$. 
In general, the chiral primary 2- and 3-point functions are non-trivial functions of the coupling constants. 
In order to discuss the coupling constant dependence of correlators we have to address issues related to 
operator mixing. This mixing is an intrinsic property of the theory,  similar to the (in general, non-abelian) 
Berry phase, which appears in perturbation theory in Quantum Mechanics\footnote{In fact, by considering 
the state-operator map, it becomes possible to relate more precisely the connection on the space of operators to the Berry 
phase of quantum states of the CFT on $S^3 \times$ time.}. The operator mixing in conformal 
perturbation theory has been discussed in several earlier works, here we mention those that are most 
relevant for our approach \cite{Seiberg:1988pf, Kutasov:1988xb, Strominger:1990pd, Bershadsky:1993cx , Ranganathan:1993vj,deBoer:2008ss, Papadodimas:2009eu}.

In order to describe the operator mixing, it is useful to think of local operators
as being associated to vector bundles over the conformal manifold. These bundles are equipped with a natural 
connection that we denote by $(\nabla_\mu)_K^L = \delta_K^L \partial_\mu +(A_\mu)_K^L$. This connection 
encodes the mixing of operators with the same quantum numbers under conformal perturbation theory. The 
curvature of this connection can be defined in terms of an integrated  4-point function
in conformal perturbation theory, by the expression
\be
\label{curvature4}
(F_{\mu\nu})_K^L \equiv [\nabla_\mu, \nabla_\nu]_K^L = \frac{1}{(2\pi)^2} \int d^4 x\, d^4 y \,\,\langle \phi^L(\infty)\, 
{\cal O}_{[\mu}(x)\, {\cal O}_{\nu]}(y)\,\phi_K(0)\rangle
~.
\ee
The index $L$ is raised with the inverse of the matrix of 
2-point functions. The reason that the RHS is not identically zero, despite the antisymmetrization in the indices $\mu,\nu$, is that the 
integral on the RHS has to be regularized to remove divergences from coincident points. The need for 
regularization is one way to understand why we end up with nontrivial operator mixing. A very thorough explanation of the 
regularization procedure needed to do the double integral is given in \cite{Ranganathan:1993vj}\footnote{In 
\cite{Ranganathan:1993vj} only 2d CFTs are discussed but several of their statements can be
generalized to 4d conformal perturbation theory.}.

In the case of ${\cal N}=2$ SCFTs, and when considering operators in the chiral ring, this double integral can be 
dramatically simplified, given that the marginal operators are descendants of chiral primaries of the form 
${\cal O}_i = \qq^4 \cdot \phi_i$ and similarly for the antiholomorphic deformations. As was shown in 
\cite{Papadodimas:2009eu}, we can use the superconformal Ward identities to move the supercharges from one 
insertion to the other, and using the SUSY algebra 
$\{\qq^i_{\alpha},\overline{\qq}^j_{\dot{\beta}}\} = 2  P_{\alpha\dot{\beta}}\delta^{ij}$ repeatedly, we get derivatives 
inside the integral. Then, by integrations by parts the integral simplifies drastically, and only picks up contributions 
which are determined by chiral ring 2- and 3-point functions and the CFT central charge $c$. The interested 
reader should consult \cite{Papadodimas:2009eu} for details. The final result is that in ${\cal N}=2$ SCFTs the 
curvature of bundles of chiral primaries is given by
\begin{subequations}
\label{ttgeneral}
\begin{align}
\label{ttgenerala}
  & [\nabla_i, \nabla_j ]_K^L = [\overline{\nabla}_i, \overline{\nabla}_j]_K^L = 0~, \\
  \label{ttgeneralb}
  & [\nabla_i , \overline{\nabla}_j]_K^L = - [C_i,\overline{C}_j]_K^L + g_{i\overline{j}}\delta_K^L \left(1+\frac{R}{4c}
  \right)~.
\end{align}
\end{subequations}
The equations on the first line express the fact that the bundles of chiral primaries are (at least locally\footnote{From now on, whenever we say `holomorphic bundle', `holomorphic section', `holomorphic function' these
terms should be understood in the sense of `locally holomorphic', since
the equations we derived are local and we have not analyzed global issues. There may be obstructions in extending the holomorphic dependence globally.}) holomorphic vector bundles
over the conformal manifold.

In the second line, $R$ is the $U(1)_R$ charge of the bundle,  $c$ the central charge of the CFT and $g_{i\overline{j}}$ is the 2-
point function of chiral primaries of $\Delta=2$, whose descendants are the marginal operators 
\eqref{marginalfromchiral}. These equations are the analogue of the \tts equations derived in 
\cite{Cecotti:1991me} for the Berry phase of the Ramond ground states and the chiral ring of
${\cal N}=(2,2)$ theories in two dimensions.

Moreover, it can be shown \cite{Papadodimas:2009eu} that the OPE coefficients of chiral primaries are covariantly 
holomorphic
\be
\label{hol3}
\overline{\nabla}_j C^I_{JK} = 0
\ee
and that OPE coefficients obey the analogue of the WDVV equations 
\cite{Witten:1989ig, Dijkgraaf:1990dj, Dijkgraaf:1990qw} which have the form
\be
\label{wdvv}
\nabla_i C_{jK}^L = \nabla_j C_{iK}^L
~.
\ee
Here, and according to our notation, the indices $i,j$ run over the marginal deformations, while $K,L$, can be any 
chiral primary. 

Finally, the supercharges and supercurrents are associated to a holomorphic line bundle ${\cal L}$ over the conformal 
manifold, whose curvature is given by\footnote{This can be shown \cite{Papadodimas:2009eu} by considering the 
general formula \eqref{curvature4} and applying it to the case where the operators $\phi_K, \phi^L$ are the 
supercurrents. Since $[{\rm supercharge}]= \int d^3x\, [{\rm supercurrent}]_0$ it is clear that the holonomy (phase) 
that the supercharges pick up under conformal perturbation theory is the same as that of the supercurrents.}

\begin{equation}
\label{curvsuper}
\begin{array}{c c}
& F_{ij} = F_{\overline{i} \,\overline{j}} = 0~, \\ 
& \vspace{-0.3cm}\\
& F_{i\overline{j}} = \frac{1}{4c} g_{i\overline{j}}~. \\
\end{array}
\end{equation}
The bundle ${\cal L}$ encodes the ambiguity of redefining the phases of the supercharges
as $\qq_\alpha^i \rightarrow e^{i\theta} \qq_\alpha^i$ and 
$\overline{\qq}_{\dot{\alpha}}^i \rightarrow e^{-i \theta} \overline{\qq}_{\dot{\alpha}}^i$ (the superconformal 
generators transform as $S\rightarrow e^{-i\theta} S$ and $\bar{S}\rightarrow e^{i\theta} \bar{S}$, while the bosonic 
generators remain invariant). It is clear that this transformation is an automorphism of the ${\cal N}=2$ 
superconformal algebra. The equations \eqref{curvsuper} are saying that in the natural connection defined by 
conformal perturbation theory, the choice of this phase varies as we move on the conformal manifold. As we see 
from \eqref{curvsuper} the curvature of the corresponding bundle ${\cal L}$ is proportional to the K\"ahler form 
of the Zamolodchikov metric.

The statements above are covariant in the sense that they hold independent of how we select the 
normalization/basis of chiral primaries as a function of the coupling constants. However, it is more practical 
to select a particular scheme, where we will see that the equations above reduce to standard partial differential 
equations for the 2- and 3-point functions, without any reference to the connection $A$ on the bundles.

A natural choice would be to select a basis of chiral primaries over the conformal manifold that consists of 
holomorphic sections of the corresponding bundles. Furthermore, from \eqref{ttgenerala} we see that it is 
possible to go to a holomorphic gauge $(A_{\overline{j}})_K^L=0$, where 
$\nabla_{\overline{j}} = \partial_{\overline{j}}$. In this gauge, the condition \eqref{hol3} simply becomes 
$\partial_{\overline{j}} C^I_{JK} = 0$, so the OPE coefficients are holomorphic functions of the couplings. 
Let us denote the chiral primaries in the gauge where they are holomorphic sections as 
$\phi_I'$ and the corresponding 2-point functions  as $\langle \phi_I' \overline{\phi}_J'\rangle = g_{I\overline{J}}'$. 
In terms of these holomorphic sections, the curvature of the underlying holomorphic bundles can be simply 
expressed as
\begin{equation}
\label{curveholsec}
 [\nabla_i , \overline{\nabla}_j]_K^L = -\partial_{\overline{j}}( {g'}^{\overline{M}L} \partial_i g'_{K \overline{M}})~,
\end{equation}
and there is no longer any explicit dependence on the connection $A$. Here we used the compatibility of the 
connection and the metric on the bundle, see \cite{Ranganathan:1993vj} for explanations.

We could continue working with these holomorphic sections, but we need to pay attention to the following 
technical detail. The marginal operators $\OO_i$ can be related to the chiral primaries $\phi'_i$ with $\Delta =2$ 
by an expression of the form ${\cal O}_i = {\qq'}^4 \cdot \phi_i'$.  
The supercharges $\qq'$ can be viewed as sections of the holomorphic bundle $\LL$ mentioned in 
equations \eqref{curvsuper}. Having chosen a convention for $\OO_i$ and $\phi'_i$ we have also chosen
the conventions for the section $\qq'$. Assuming $\OO_i$ is holomorphic (from \eqref{actiondeformation}),
the above choice of the holomorphic section $\phi'_i$ implies that $\qq'$ is a holomorphic section of $\LL$.
These conventions for the supercharges are not the standard ones following from the supersymmetry algebra.
In the standard conventions, although the overall phase of the supercharges can be redefined in a coupling-constant dependent way due to the $U(1)$ automorphism of the algebra, the
``magnitude'' of the normalization of the supercharges is fixed in order to satisfy the standard supersymmetry algebra 
$\{\qq^i_{\alpha},\overline{\qq}^j_{\dot{\beta}}\} = 2  P_{\alpha\dot{\beta}}\delta^{ij}$. 
Equivalently, the normalization of the 2-point function of the corresponding supercurrents
is independent of the coupling constant. Since the supercharges $\qq$ with this standard choice have 
constant magnitude, they cannot be a holomorphic section of the bundle $\LL$.\footnote{Had they been 
holomorphic sections with constant magnitude, we would conclude from \eqref{curveholsec} that the 
curvature of ${\cal L}$ is zero, which is inconsistent with the direct computation leading to \eqref{curvsuper}.}
Hence, the standard $\qq$ and the $\qq'$ above are different types of sections. 
What is the precise relation between them?

Equation \eqref{curvsuper} implies that the combination 
\beq
\label{QQrelation}
\qq' = e^{\frac{{\cal K}}{c'}} \qq~, ~~ c' = 8\times 192 \times c
\eeq
can be a holomorphic section for an appropriate choice of the (coupling-constant dependent) phase of $\qq$. 
${\cal K}$ is the K\"ahler potential of the Zamolodchikov metric. Notice that the appropriate choice of the 
phase of $\qq$ 
depends on the choice of K\"ahler gauge. Under a K\"ahler transformation, 
${\cal K} \rightarrow {\cal K} + f + \bar{f}$ (where $f$ ($\bar f$) is (anti)holomorphic), the section 
$\qq'$ in \eqref{QQrelation} becomes
$$
e^{\frac{2f}{c'}} \,
e^{i \frac{2 {\rm Im} f}{c'}}
\, \qq'
~.$$
There is an overall holomorphic factor $e^{\frac{2f}{c'}}$ and the original phase of $\qq$ has been shifted.
With these specifications \eqref{QQrelation} is the relation between $\qq$ and $\qq'$ that we are looking for.

This suggests the following choice of conventions: select chiral primaries $\phi_I$ at any level of 
$R$-charge $R$ so that $\phi'_I = e^{-{\frac{R}{c'}}{\cal K}} \phi_I$ are holomorphic 
sections. Equivalently, if we have already a choice of holomorphic sections $\phi_I'$ (as above), then
we define a new non-holomorphic basis by 
$\phi_I =e^{{\frac{R}{c'}}{\cal K}} \phi'_I$.\footnote{Again, this definition of $\phi_I$ depends 
on the K\"ahler gauge and the resulting 2-point function $g_{I\overline{J}}$ transforms as 
$g_{I\overline{J}} \rightarrow e^{\frac{2R}{c'}(f+\bar f)} g_{I\overline{J}}$ under K\"ahler transformations. 
Happily, this dependence drops out of the 
final equation \eqref{maintt}, which is indeed invariant under K\"ahler transformations.
We are grateful to M. Buican, for discussions which led us to an investigation
of the invariance of our statements under K\"ahler transformations.} 
The corresponding 2-point functions obey the relation 
$g_{I\overline{J}} =e^{\frac{2R}{c'}{\cal K}} g'_{I \overline{J}}$. 
This choice ensures that ${\cal O}_i={\qq'}^4\cdot \phi'_i = \qq^4 \cdot \phi_i$, where $\qq$ are supercharges with
the standard normalization.
The non-holomorphicity of $\phi_i$ precisely cancels the non-holomorphicity of $\qq$. 
In addition, the general OPE coefficients are the same in the two bases, $C^I_{JK} = C^{I'}_{J'K'}$, as a
consequence of $R$-charge conservation. 

In the $\phi_I$-basis the curvature of the bundles becomes 
\begin{align}
[\nabla_i , \overline{\nabla}_j]_K^L =- \partial_{\overline{j}}( {g'}^{\overline{M}L} \partial_i g'_{K \overline{M}}) = -
\partial_{\overline{j}}( g^{\overline{M}L} \partial_i g_{K \overline{M}})  + \frac{R}{4c} g_{i\overline{j}} \delta_K^L
~.
\end{align}
Inserting into \eqref{ttgeneralb} we obtain the partial differential equations\footnote{The reader familiar with the 2d 
\tts equations should notice that the last term $- g_{i\overline{j}}\delta_K^L$ can be effectively removed by a slight 
redefinition, see the discussion around $\eqref{SU(2)aea}$ for an example.}
\begin{align}
\partial_{\overline{j}}( g^{\overline{M}L} \partial_i g_{K \overline{M}})   =  [C_i,\overline{C}_j]_K^L 
- g_{i\overline{j}}\delta_K^L 
~.
\end{align}

\subsection{Differential equations for 2- and 3-point functions of chiral primaries}
\label{sec:2and3ptf}

The result of this choice of gauge (scheme) is that the \tts equations reduce to differential equations for the 
2- and 3-point functions, where there is no explicit appearance of the connection on the bundles. For the sake 
of clarity we summarize here the detailed form of the equations with all indices written out
\be
\label{maintt}
\frac{\partial}{\partial\overline{\lambda^j}} \left( g^{\overline{M}L}\, \frac{\partial}{\partial \lambda^i}\, g_{K\overline{M}}\right)  = C_{iK}^P\, g_{P\overline{Q}}\, C^{* \overline{Q}}_
{\overline{j}\overline{R}} \,
g^{\overline{R}L} - g_{K\overline{N}} \,C^{*\overline{N}}_{\overline{j} \overline{U}}\, g^{\overline{U} V}\, C_{iV}^L  -
g_{i\overline{j}} \,\delta_K^L ~.
\ee
As we can see these differential equations relate the coupling constant dependence of 2- and 3-point functions of 
various chiral primaries. They have to be supplemented by equation \eqref{hol3}, which in this gauge takes the 
simpler form
\be
\label{mainttB}
\frac{\partial}{\partial \overline{\lambda}^j} C_{IJ}^K = 0
~,
\ee
and the WDVV equations \eqref{wdvv}
\beq
\label{wdvvholg}
 {\partial C_{j K}^L \over \partial \lambda^i} -{\partial C_{i K}^L \over\partial  \lambda^j} 
= 
g^{\overline Q L}\, \partial_i g_{P\overline Q}\, C_{jK}^P
- C_{jP}^L \, g^{\overline{Q} P} \, \partial_i g_{K\overline{Q}}
-(i \leftrightarrow j)
~.
\eeq
In the examples that we will study later the conformal manifold
is 1-(complex) dimensional, hence the WDVV equations are trivially obeyed and that is why we do not discuss them 
any further. In other ${\cal N}=2$ theories with higher dimensional conformal manifolds they may be nontrivial. 

Let us elaborate a little further on the notation in equation \eqref{maintt}. The lowercase indices $i,\overline{j}$ 
run over (anti)-chiral primaries of $\Delta=2, R = \pm 4$, or equivalently, over the marginal directions along the 
conformal manifold. We remind that chiral primaries of $R=\pm4$ and dimension 
$\Delta =2$ are those  whose descendants are the marginal operators corresponding to 
$\lambda^i , \lambda^j$ on the LHS. The capital indices run over general chiral primaries of any R-charge. 
These equations can be applied for each possible sector of chiral primaries. 
The function $g_{K\overline{M}}$ is the 2-point function of chiral primaries of charge $R$. The OPE coefficients 
$C_{iK}^P$ relate the chiral primaries of charge $R$ (corresponding
to the index $K$) to the chiral primaries of charge $R+4$ (corresponding to the index $P$). 
The indices $U,V$ correspond to chiral primaries of charge $R-4$. Finally by 
$C^{*\overline{Q}}_{\overline{j} \overline{R}}$ we mean $(C^Q_{jR})^*$.

\subsection*{Remark on the curvature of the Zamolodchikov metric}

If we consider equation \eqref{maintt} specifically for the bundle of chiral primaries of R-charge $4$ (whose 
descendants are the marginal operators) and using \eqref{f192} and the general formula for the Riemann 
tensor of a K\"ahler manifold we get the equation
\be
\label{riemannmod}
R_{i \overline{j} k}^l = - C_{ik}^M g_{M\overline{N}} C^{* \overline{N}}_{\overline{j} \overline{q}}g^{\overline{q}l} 
+ g_{k\overline{j}}\delta_i^l + g_{i\overline{j}}\delta_k^l
\ee
We notice that the curvature of the conformal manifold obeys an equation, which is reminiscent of the one for the 
moduli space of 2d ${\cal N}=(2,2)$ SCFTs with general values of the central charge, as 
some sort of generalization of special geometry \cite{Strominger:1990pd, Bershadsky:1993cx}.

\subsection*{Note on normalization conventions}

We emphasize once again that the differential equations \eqref{maintt} hold in a particular choice of 
normalization conventions described near the end of section \ref{ttmaineqs}. The benefit of this choice is that it 
allows us to circumvent the details of a non-trivial connection on the chiral primary bundles. These normalization
conventions are typically different from the more common ones in conformal field theory where one
diagonalizes the 2-point functions of conformal primary fields,
\beq
\label{normalizeaa}
\left< \phi_K(x) \overline{ \phi}_{L}(0) \right> = \frac{\delta_{K \overline L}}{|x|^{2 \Delta}}
~.
\eeq
In the conventions \eqref{normalizeaa} the OPE coefficients $C_{IJ}^K$ are no longer holomorphic functions of the
marginal couplings and therefore do not obey \eqref{mainttB} (but they still obey \eqref{hol3}).

In the examples of section \ref{sqcd} a natural basis of chiral primaries will lead to the holomorphic
gauge of equation \eqref{mainttB}. Once there is a solution of the \tts equations in this basis, it is not hard to 
rotate to the more conventional basis \eqref{normalizeaa}.

\subsection{Global issues}

When studying the equations \eqref{maintt} it is important and interesting to explore certain global 
issues\footnote{We are grateful to M. Buican for discussions on this.} of the bundles of chiral primaries over the 
conformal manifold ${\cal M}$. The equations are {\it local}, since they were derived in conformal perturbation 
theory, but the conformal manifold may have special points (e.g. the weak coupling point $g_{YM}=0$) and 
nontrivial topology like in the class $\SS$ theories \cite{Gaiotto:2009we,Alday:2009aq}, where the conformal 
manifold is related to the moduli space of punctured Riemann surfaces. Because of these global issues, it is 
conceivable that in certain theories, the connection on the space of operators is not entirely determined by the local 
curvature expression \eqref{maintt}, but there may be additional ``Wilson line''-like configurations around the 
special points/nontrivial cycles on the conformal manifold. Moreover, whether we can find global holomorphic 
sections or not and if we can set $\overline{\partial} C = 0$ globally, may be a nontrivial question. In this paper, 
since we are dealing mostly with the simpler superconformal QCD theories, we will not go into these global issues 
but we are planning to return to them in future work.

\subsection{Solving the \tts equations}

The resulting equations \eqref{maintt} are a set of coupled differential equations for the 2- and 3- point functions of 
chiral primaries. In certain 2d ${\cal N}=(2,2)$ QFTs the \tts equations could be solved \cite{Cecotti:1991me, Cecotti:
1991vb} just from the requirement that the 2-point functions must be positive and from knowing the correlators in 
the weak coupling region. For this to work it was important that the chiral ring in 2d is finite dimensional. For 
example, in ${\cal N}=(2,2)$ SCFTs a unitarity bound constrains the R-charge by $|q|\leq \frac{c}{3}$, 
which shows that in theories with reasonable spectrum the chiral ring is truncated. In 4d ${\cal N}=2$ SCFTs the 
chiral ring has no known upper bound in R-charge and if we try to apply these equations we end up with an infinite 
set of coupled differential equations. For instance, while in certain 2d examples one gets equations corresponding 
to the periodic Toda chain \cite{Cecotti:1991vb}, in 4d ${\cal N}=2$ SCFTs we find equations similar to the
semi-infinite Toda-chain (this will become more clear in section 4). Unlike what happened to 2d examples 
\cite{Cecotti:1991me, Cecotti:1991vb}, we have not been able to find a way to uniquely determine a solution of 
these equations, just from the requirement of positivity of the 2-point functions and the boundary conditions at 
weak coupling. 

On the other hand, in certain 4d ${\cal N}=2$ SCFTs, these equations have a recursive structure: if we somehow fix 
the coupling constant dependence of the lowest nontrivial chiral primaries, then the equations predict the 2- and 3-
point functions of higher-charge chiral primaries. As we explained in section 2, the 2-point functions of chiral 
primaries of R-charge 4, are proportional to the Zamolodchikov metric on the conformal manifold.

Hence, if we knew the exact Zamolodchikov metric as a function of the coupling, we would also know the 2-point 
function of chiral primaries of R-charge $4$, and then by plugging this into the sequence of \tts equations we would 
be able to compute the 2- and 3-point functions of an infinite number of other chiral primaries. Progress in this 
direction becomes possible after the recent proposal \cite{Gerchkovitz:2014gta}, which relates the partition function 
of ${\cal N}=2$ SCFTs on $S^4$ computed by localization in the work of Pestun \cite{Pestun:2007rz}, 
to the K\"ahler potential of the Zamolodchikov metric on the moduli space.

While this strategy allows us to partly solve the \tts equations, it would be interesting to explore whether it is 
possibile to determine the relevant solution of these equations without input from localization. This could perhaps 
be possible by demanding positivity of all 2-point functions of chiral primaries over the conformal manifold 
supplemented by some weak coupling perturbative data, in analogy to what was done in \cite{Cecotti:1991vb}. This 
is a very speculative possibility, which if true, would in principle lead to an alternative computation of the nontrivial 
information encoded in the sphere partition function, without the use of localization. We plan to investigate this 
further in future work.

\subsection{Extremal correlators}
\label{extremal}

By computing the 2- and 3-point functions of chiral primaries we can also get exact results for more general 
``extremal correlators''. These are correlators of the form
\begin{align}
\langle \phi_{I_1}(x_1)...\phi_{I_n}(x_n) \overline{\phi}_{J}(y)\rangle~,
\end{align}
where $\phi_{I_k}$ are chiral primaries and $\overline{\phi}_{J}$ is antichiral, with $R$-charges related as 
$R_J =- \sum_k R_{I_k}$. 

First, it is convenient to use a conformal transformation of the form
\begin{equation}
 \label{confinv}
x^{\mu'} = \frac{x^\mu - y^\mu}{|x-y|^2}
\end{equation}
to write the correlator as
\be
\label{extgen}
\langle \phi_{I_1}(x_1)...\phi_{I_n}(x_n) \overline{\phi}_{J}(y)\rangle = \frac{
\langle \phi_{I_1}(x_1')...\phi_{I_n}(x_n') \overline{\phi}_{J}(\infty)\rangle}
{|x_1-y|^{2\Delta_1}\ldots |x_n-y|^{2\Delta_n}}~,
\ee
where the $x'$'s on the RHS are related to $x$'s by \eqref{confinv}.

For an extremal correlator in ${\cal N}=2$ SCFT, the superconformal Ward identities imply that
\be
\label{extinf}
\langle \phi_{I_1}(x_1)...\phi_{I_n}(x_n) \overline{\phi}_{J}(\infty)\rangle
\ee
is independent of the positions $x_i$. Consequently, we are free to evaluate it in any particular limit. Let us define a 
new chiral primary $\phi_I$ by fusing together all the chiral primaries
\begin{equation}
\label{fusionext}
\phi_I(0) \equiv \lim_{\{x_i\} \rightarrow 0} \phi_{I_1}(x_1)\times...\times \phi_{I_n}(x_n)~,
\end{equation}
where the symbol $\times$ refers to an OPE. Notice that, since all operators are chiral primaries, this multi-OPE is 
non-singular and associative, so the limit is well defined and it is simply given by a chiral primary $\phi_I$ of 
charge $R_I = \sum_k R_{I_k}$. Then we find that
\begin{equation}
\label{2pointext}
\langle \phi_{I_1}(x_1)...\phi_{I_n}(x_n) \overline{\phi}_{J}(\infty)\rangle = \langle \phi_I(0) \overline{\phi}_J(\infty)
\rangle = g_{I \overline{J}}~,
\end{equation}
where on the last step we got the usual 2-point functions of chiral primaries \eqref{def2point}. Due to the 
associativity of the chiral ring we can also write
\begin{equation}
\label{assoccr}
g_{I\overline{J}} = C_{I_1 I_2}^{M_1}\, C_{ M_1 I_3}^{M_2}\,... \,C_{M_{n-2} I_n }^{M_{n-1}} \,
g_{M_{n-1} \overline{J}}
\end{equation}
Re-instating the full coordinate dependence from \eqref{extgen}, we can write the following formula for extremal
correlators
\begin{equation}
\label{extermalcor}
\langle \phi_{I_1}(x_1)...\phi_{I_n}(x_n) \overline{\phi}_{J}(y)\rangle = \frac{g_{I \overline{J}}}
{|x_1-y|^{2\Delta_1}\ldots |x_n-y|^{2\Delta_n}}~.
\end{equation}
So according to our argument, extremal correlators can be uniquely determined by the chiral ring 2- and 3-point 
functions, which were used in formulae \eqref{fusionext} (OPE coefficients) and \eqref{2pointext} (2-point functions).

\subsection{${\cal N}=4$ theories}
\label{N4}

Until this point we considered general theories with $\NN=2$ supersymmetry. 
It is interesting to ask parenthetically how the formalism captures the properties of $\NN=4$ theories.
An ${\cal N}=4$ theory can also be written as an ${\cal N}=2$ theory, so our formalism should apply. 
The R-symmetry  $SU(2)_R\times U(1)_R$ of the ${\cal N}=2$ viewpoint, is embedded inside the underlying  
$SO(6)_R$ of the full ${\cal N}=4$ theory. We proceed to flesh out the pertinent details and verify that the \tts 
equations work correctly in ${\cal N}=4$ theories. 

Consider an ${\cal N}=4$ gauge theory with semi-simple gauge group ${\cal G}$. The theory has 6 real scalars 
$\Phi^{\rm I}, {\rm I} =1,...,6$. It is useful to define the complex combination
\be
\varphi = \Phi^1 + i \Phi^2 
\ee
which is the bottom component of an $SU(3)$ highest weight $\NN=1$ superfield. 
The $U(1)_R$ symmetry that rotates this field corresponds to rotations on the 1-2 plane. 
The chiral primary, whose descendant is the $\NN=4$ marginal operator, has the form
\be
\phi_2 \propto \Tr[\varphi^2] 
~.
\ee
From the ${\cal N}=4$ viewpoint this is the superconformal primary of the ${1\over 2}$-BPS short 
representation of ${\cal N}=4$ which contains, among other operators, the R-symmetry currents, 
stress tensor and marginal operators.

General chiral primaries of charge $R$ in ${1\over 2}$-BPS representations can be deduced from multitrace 
operators of the form
\be
\phi_K \propto \Tr[\varphi^{n_1}]...\Tr[\varphi^{n_k}] ,
\ee
where $2\sum n_i = R$. The trace is taken in the adjoint of ${\cal G}$. 

The conformal manifold of this theory is parametrized by the complexified coupling
\be
\label{sqcdaa}
\tau = {\theta \over 2\pi} +  {4\pi i \over g_{YM}^2}
\ee
up to global identifications due to S-duality transformations. 
$\theta$ denotes the $\theta$-angle and $g_{YM}$ the Yang-Mills coupling.
An important point is that for ${\cal N}=4$ theories the Zamolodchikov metric on the 
conformal manifold does not receive any quantum corrections and in our conventions is equal to
\be
\label{zmetricn4}
G_{\tau \bar{\tau}}^{{\cal N}=4} = 96 {c \over {\rm Im}\tau^2}
~.
\ee
This means that the conformal manifold is locally a two-dimensional homogeneous space of constant negative 
curvature. The marginal operators ${\cal O}_\tau, \overline{\cal O}_{\tau}$ can be thought of as holomorphic and 
antiholomorphic tangent vectors to the conformal manifold. Since the manifold \eqref{zmetricn4} has nonzero 
curvature, the marginal operators have a nontrivial connection.

On the other hand, we will argue that the bundles encoding the connection for chiral primaries have vanishing 
curvature in ${\cal N}=4$ theories. This can be seen as follows: while from the ${\cal N}=2$ 
point of view the chiral primaries are only charged under $U(1)_R$, in the underlying ${\cal N}=4$ theory they 
belong to representations of $SO(6)_R$. Since the conformal manifold is one-complex dimensional and the 
holonomy of the tangent bundle is only $U(1)$, it is not possible to have notrivial $SO(6)$-valued curvature for 
bundles over the conformal manifold,  without breaking the $SO(6)$ invariance of the theory.

Hence we conclude that the bundles of chiral primaries for ${\cal N}=4$ theories must have vanishing curvature. 
One might wonder, how this statement can be consistent with the fact that the tangent bundle has nontrivial 
curvature and the fact that the marginal operators are descendants of the chiral primaries. The resolution is simple. 
Recalling the relation ${\cal O}_\tau = \qq^4\cdot \phi_2$, we can see that the curvature corresponding 
to ${\cal O}_\tau$ is given by the sum of the curvature of the supercurrents plus that of $\phi_2$. Since the latter is 
vanishing, we learn that the curvature of the tangent bundle comes entirely from that of the supercharges 
\eqref{curvsuper}. It is easy to check that, using \eqref{curvsuper}, the relation ${\cal O}_\tau = \qq^4\cdot \phi_2$ and 
comparing with the curvature of the tangent bundle of \eqref{zmetricn4}, all factors work out correctly. 

Alternatively, we can verify the fact that the chiral primaries in ${\cal N}=4$ have vanishing curvature 
directly from the \tts equations. This can be done in two steps. The first step is to observe that in ${\cal N}=4$ 
theories, we have a non-renormalization theorem for 3-point functions \cite{Lee:1998bxa,D'Hoker:1998tz,D'Hoker:1999ea,Intriligator:1998ig,Intriligator:1999ff,Eden:1999gh,Petkou:1999fv,Howe:1999hz,Heslop:2001gp,Baggio:2012rr}, 
which can be expressed in equations as
\be
\label{nablaC}
\nabla_\tau C = \overline{\nabla}_{\overline{\tau}} C= 0
~.
\ee
The second step requires taking the covariant derivative (either $\nabla$ or $\overline{\nabla}$) of both sides of the 
\tts equation \eqref{ttgeneralb}. The covariant derivative of the RHS, which involves the two-point function 
coefficients $g$ and the 3-point function coefficients $C$, vanishes from \eqref{nablaC} and the compatibility of $g$ 
with the connection, which implies $\nabla g = \overline{\nabla} g =  0$. The vanishing of the covariant derivative of 
the RHS implies that the covariant derivative of the LHS also vanishes, from which we deduce that the bundles 
must have covariantly constant curvature. This allows a direct evaluation of the curvature in the weak coupling limit. 
Hence, in order to show that the curvature vanishes in $\NN=4$ theories for all values of the coupling, it is enough 
to show that the RHS of the \tts equations \eqref{ttgeneralb} vanishes  in the weak coupling limit. 

All ingredients on the RHS of \eqref{ttgeneralb} can be evaluated --- in principle --- by standard, alas rather involved 
in general, Wick contractions. In appendix \ref{LemmaApp} we provide an alternative derivation of the following 
general combinatoric/group theoretic identity
\be
\label{grouplemma}
\left\{ 
-[C_2,\overline{C}_{\overline{2}}]_K^L + g_{2 \overline{2}} \delta_K^L \left(1 + {R \over {\rm dim}\,{\cal G}}  \right)
\right\}_{\rm tree}=0 
~.
\ee
This is an identity\footnote{It is quite possible that this equation corresponds to a natural group-theoretic statement, 
but we have not yet investigated this in detail. See also section \ref{pertchecks} for related explicit tree-level 
2-point functions.} for free-field contractions between traces that should hold for any semi-simple group ${\cal G}$. 
The subscript 2 refers to the chiral primary $\phi_2 = \Tr [\varphi^2]$.

Using this identity, we can demonstrate the desired result, i.e. that the RHS of the \tts equation vanishes for 
${\cal N}=4$ theories: in standard ${\cal N}=4$ gauge theories the central charge is related to ${\rm dim} {\cal G}$ 
by
$$
c = {{\rm dim }{\cal G} \over 4}
~.
$$
Inserting this formula into \eqref{grouplemma} we find
\be
\label{curvn4f}
-[C_2,\overline{C}_{\overline{2}}]_K^L + g_{2 \overline{2}} \delta_K^L \left(1 + {R \over 4 c}  \right) =0
\ee
which is precisely what we wanted to show.

As a final comment we would like to clarify a possibly confusing point. The \tts equations \eqref{maintt} predict that 
the chiral primaries in ${\cal N}=2$ theories have nonzero curvature even in the limit of weak coupling. Indeed, 
the relation between $c$ and ${\rm dim} {\cal G}$ is different for ${\cal N}=2$ theories compared to 
${\cal N}=4$ theories and as a result \eqref{curvn4f} does not hold in ${\cal N}=2$ theories (\eqref{grouplemma}, 
however, does hold). On the other hand, we argued that the curvature of operators in conformal 
perturbation theory is computed by \eqref{curvature4}. In the free limit the 4-point function inside the double 
integral, relevant for the computation of the curvature of chiral primaries, is the same in ${\cal N}=2$ and 
${\cal N}=4$ theories. How can it then be, that in ${\cal N}=2$ the bundle of primaries has 
nonzero curvature even in the weak coupling limit, while in ${\cal N}=4$ the curvature vanishes? 

The answer is that the two processes, of taking the zero coupling limit and of doing the double regularized integral, do not commute. In principle, the correct computation is to first compute the integral at some finite value of the coupling, and then send the coupling to zero. If one (wrongly) first takes the zero coupling limit inside the integral, then operators whose conformal dimension takes ``accidentally'' small value at zero coupling, start to contribute to the double integral. At infinitesimally small coupling these operators lift and their contribution discontinously drops out of the double integral. Such operators are different between ${\cal N}=2$ and ${\cal N}=4$, thus resolving the aforementioned paradox.

\section{$\NN=2$ superconformal QCD as an instructive example}
\label{sqcd}

\subsection{Definitions}

The $\NN=2$ SYM theory with gauge group $SU(N)$ coupled to $2N$ hypermultiplets 
(in short, $\NN=2$ superconformal QCD or SCQCD) is a well known superconformal field theory for any value of 
the complexified gauge coupling constant \eqref{sqcdaa}. This theory will serve as a testing ground for the general 
ideas presented above. The bosonic field content of the theory comprises of: $(a)$ the gauge field $A_\mu$ 
and a complex scalar field $\varphi$ in the adjoint representation of the gauge group (both are part of the $\NN=2$ 
vector multiplet), and $(b)$ $2N$ doublets of complex scalars $Q_{\II}$ $(\II=\pm)$ in the fundamental 
representation of the gauge group, that belong to $2N$ $\NN=2$ hypermultiplets. The global symmetry group is 
$U(2N) \times SU(2)_R \times U(1)_R$. $U(2N)$ is a flavor symmetry rotating the hypermultiplets and 
$SU(2)_R \times U(1)_R$ is the $\NN=2$ R-symmetry. More details about the theory are summarized in appendix 
\ref{conventions}.

The generators of the $\NN=2$ chiral ring, as defined in section \ref{chiralring}, are the 
single-trace superconformal primaries
\beq
\label{sqcdab}
\phi_{\ell} \propto \Tr\, [ \varphi^{\ell} ] ~, ~~ \ell=2,3,\ldots,N
~.
\eeq
The proportionality constant is convention-dependent (specific convention choices will be made below).
The remaining fields of the chiral ring are generated by products of the fields \eqref{sqcdab};
in the weak-coupling formulation of the theory chiral primaries with $\ell>N$ are related 
to the primaries with $\ell \leq N$ by polynomial equations dictated by the Cayley-Hamilton theorem of
$N\times N$ matrices.

$\NN=2$ superconformal QCD has a single (complex) exactly marginal deformation \eqref{actiondeformation} with 
coupling $\tau$ \eqref{sqcdaa}. The exactly marginal operator $\OO_\tau$ is a descendant of the chiral primary 
field $\phi_2$ 
\beq
\label{sqcdaba}
\OO_\tau = {\cal \qq}^4 \cdot \phi_2
~.\eeq

We note in passing that the chiral ring defined in terms of an $\NN=1$ subalgebra contains the additional 
mesonic superconformal primaries  
\beq
\label{sqcdabb}
\MM_{3\, \II}^{~\JJ} \propto \left( Q_{\II\, j} {\overline Q}^{\JJ\, j} \right) 
-\frac{1}{2} \left( Q_{\KK\, j} {\overline Q}^{\KK \, j} \right) ~ \delta_{\II}^{\JJ} 
~.
\eeq
A sum over the gauge group indices is implicit, the index $j=1,\ldots,2N$ runs over the number of
hypermultiplets, $\II,\JJ,\KK=\pm$ are $SU(2)_R$ indices, and the subindex $\bf 3$ denotes that this particular 
combination belongs in a triplet representation of the $SU(2)_R$\footnote{For a complete analysis of the 
shortening conditions of the $\NN=2$ superconformal algebra in general theories we refer the reader to 
\cite{Dolan:2002zh}. For an application to the $\NN=2$ superconformal QCD theories see for example 
\cite{Gadde:2009dj}.}. Such primaries are not part of the $\NN=2$ chiral ring defined in section \ref{chiralring}
and therefore will not be part of our analysis.

\subsection{$SU(2)$ with 4 hypermultiplets}

We begin the discussion with the $SU(2)$ case which provides a simple clear demonstration 
of the general ideas in section \ref{ttreview}. In this case, $\phi_2$ is the {\it single} chiral ring generator. 
We normalize $\phi_2$ by requiring the validity of the conventions \eqref{actiondeformation}, 
\eqref{marginalfromchiral}, \eqref{f192}
(see also section \ref{2tree} for an explicit tree-level implementation of these conventions).
We notice that since $\OO_\tau$ is, by this definition, related to a holomorphic section of the tangent bundle 
of the conformal manifold, then as explained in section \ref{ttreview}, $\phi_2 \propto \Tr[\varphi^2]$ (with a 
normalization that is a holomorphic function of $\tau$) is a {\it non-holomorphic} section of the bundle of chiral 
primaries. A holomorphic $\phi_2$ arises by multiplying $\Tr [\varphi^2]$ with the non-holomorphic factor 
$e^{-\frac{\KK}{384 \, c}}$, where $\KK$ is the K\"ahler potential for the Zamolodchikov metric.

In addition, the chiral ring includes a unique chiral primary $\phi_{2n} \propto \left(\Tr[\varphi^2]\right)^n$ 
at each scaling dimension $\Delta= 2n$ $(n\in {\mathbb Z}_+)$ (generated by $\phi_2$ with repeated 
multiplication). We normalize the higher order chiral primaries $\phi_{2n}$ $(n>1)$ by requiring the OPE
\beq
\label{SU(2)aa}
	\phi_2(x) \phi_{2n}(0)  = \phi_{2n+2}(0) + \ldots
\eeq
which fixes the OPE coefficients
\beq
\label{SU(2)ab}
	C_{2\, 2n}^{2n+2} = 1
~.
\eeq
Notice that this choice is consistent with the holomorphic gauge \eqref{mainttB}. Moreover, 
as a straightforward consequence of the associativity
of the chiral ring all the non-vanishing OPE coefficients are fixed to one; 
namely, one can further show that
\beq
\label{SU(2)ac}
C_{2n\, 2m}^{2(n+m)} =1
~.
\eeq

\subsubsection{\tts equations and exact 2- and 3-point functions}

In these conventions the 2-point functions of the chiral primaries $\phi_{2n}$ 
\beq
\label{SU(2)ad}
	\left< \phi_{2n}(x) \overline{\phi}_{2n}(0)\right> = \frac{g_{2n}(\tau,\overline{\tau})}{|x|^{4n}}
\eeq
have a non-trivial dependence on the modulus $\tau$.
Our purpose is to determine the exact form of the functions $g_{2n}(\tau,\bar\tau)$. This will immediately
provide information about 3-point functions as well.

Since we have a one-dimensional sequence of chiral primaries without any non-trivial degeneracies, 
the \tts equations \eqref{maintt} assume the following particularly simple form
\be
\label{SU(2)ae}
\partial_{\tau}\partial_{\overline \tau} \log g_{2n} = \frac{g_{2n+2}}{g_{2n}} - \frac{g_{2n}}{g_{2n-2}} - g_2
~,
\ee
where $n=1,2,...$ and $g_0 = 1$ by definition. 
This infinite sequence of differential equations can be recast as the more familiar semi-infinite Toda chain
\beq
\label{SU(2)aea}
\partial_{\tau}\partial_{\bar \tau} q_n = e^{q_{n+1}-q_n} - e^{q_n-q_{n-1}} ~, ~~
n=2,\ldots
\eeq 
by setting $g_{2n}={\rm exp}\left( q_n - \log Z_{S^4} \right)$.
A reality condition on $q_n$ implies that $g_{2n}$ are positive, which is expected by unitarity.
In section \ref{checks} we collect several perturbative checks of equations \eqref{SU(2)ae}.

It may be interesting to classify the most general solution of the equations \eqref{SU(2)ae}, 
subject to positivity over the entirety of the conformal manifold, but this is beyond the scope of the 
current paper.\footnote{We do not expect positivity alone to fix the solution uniquely. 
It is worth exploring the possibility that positivity, in combination with the data of higher order perturbative 
corrections around the point weak coupling point ${\rm Im\tau} = \infty$, might lead to a unique solution, 
in analogy to 2d examples \cite{Cecotti:1991vb}.}
Instead, in what follows we will use these equations to solve recursively for the 2-point functions as follows
\be
\label{SU(2)af}
	g_{2n+2} = g_{2n}\, \partial_{\tau}\partial_{\bar \tau} \log g_{2n} + \frac{g_{2n}^2}{g_{2n-2}} + g_2\, g_{2n},
	~~ n=1,2,\ldots
\ee
Knowledge of a single 2-point function, e.g. $g_2$, implies recursively the precise form of all the rest.
As we show now, for $SU(2)$ this provides the complete non-perturbative determination 
of the 2- and 3-point functions of all chiral primary operators.

\subsubsection*{Exact 2-point functions}

We can use supersymmetric localization on $S^4$ and the 
formula \eqref{g2fromZ} to determine the exact coupling constant dependence of $g_2$ . 
For the $SU(2)$ SCQCD theory an integral expression for the sphere partition function gives 
\cite{Pestun:2007rz}
\beq
\label{SU(2)ba}
	Z_{S^4}(\tau,\overline{\tau}) = \int_{-\infty}^\infty da\, e^{-4\pi {\rm Im}(\tau) a^2} 
	(2a)^2 \frac{H(2ia) H(-2ia)}{(H(ia) H(-ia))^4} |Z_{\rm inst}(a,\tau)|^2
~.
\eeq
$H$ is a function on the complex plane defined in terms of the Barnes $G$-function \cite{Barnes} as
\beq
\label{SU(2)baa}
H(z) = G(1+z)\, G(1-z)
~.
\eeq 
Further details are summarized for the convenience of the reader in appendix \ref{SpecialFunctions}.
$Z_{\rm  inst}$ is the Nekrasov partition function \cite{Nekrasov:2002qd} that incorporates the 
contribution from all  instanton sectors.

Consequently, implementing \eqref{g2fromZ} we obtain the exact 2-point function of the lowest chiral 
primary $\phi_2$ as
\be
\label{SU(2)bb}
g_2 = \partial_\tau \partial_{\overline{\tau}} \log Z_{S^4}
~.
\ee
The 2-point functions of the higher order chiral primaries can be computed recursively using 
\eqref{SU(2)af}. We will return to the resulting expressions momentarily.

\subsubsection*{Exact 3-point functions}

The general non-vanishing 3-point function 
\beq
\label{SU(2)caa}
\left< \phi_{2m} (x_1)\, \phi_{2n}(x_2)\, \overline{\phi}_{2m+2n} (y) \right> 
= \frac{C_{2m\, 2n\, \overline{2m+2n}}}{|x_1-y|^{4m} |x_2-y|^{4n} }
\eeq
follows immediately from the above data since
\beq
\label{SU(2)cab}
C_{2m\, 2n\, \overline{2m+2n}} = 
C_{2m\, 2n}^{2(m+n)} g_{2(m+n)} = g_{2(m+n)}
~.
\eeq
In the second equality we made use of the OPE coefficients \eqref{SU(2)ac}. 
This formula provides the non-perturbative 3-point functions of chiral primaries as  
a function of the modulus $\tau$, including all instanton corrections. Following section 
\ref{extremal} it is straightforward to extend this result to any extremal correlator of chiral primaries.

While the above normalization of the chiral primaries is very convenient for the type of computations of the 
previous section, it is common in conformal field theory to work with orthonormal fields $\hat \phi_I$ for which
\beq
\label{SU(2)ca}
	\left<\hat \phi_I (x) \overline{\hat \phi}_{J}(0)\right> = \frac{\delta_{I\bar{J}}}{|x|^{2\Delta}}
~.
\eeq
In these conventions, the OPE coefficients $\hat C_{IJ}^K$ 
depend non-trivially on the moduli. Converting to this normalization in the case at hand 
we find the structure constants
\beq
\label{SU(2)cb}
	{\hat C}_{2m \, 2n\, \overline{2m+2n}} = 
	\sqrt{\frac{g_{2m+2n}}{g_{2m}\, g_{2n}}}~.
\eeq

\subsubsection{Perturbative expressions}

The \tts equations have allowed us to obtain exact results for 2- and 3-point functions of the
chiral primary fields. The resulting expressions depend implicitly on the $S^4$ partition function of the 
$SU(2)$ theory, which is given in terms of an one-dimensional integral \eqref{SU(2)ba}.
It is interesting to work out the first few orders in the perturbative expansion of the exact expressions.
This will be useful later on in section \ref{checks} when we compare against independent computations
in perturbation theory.

\subsubsection*{0-instanton sector}

Consider the perturbative contributions around the weak coupling regime $g_{YM} \rightarrow 0$, or 
equivalently $\tau \rightarrow +i \infty$.
Working with the perturbative (0-instanton) part of the $S^4$ partition function we obtain 
\beq
\label{SU(2)da}
Z_{S^4}^{(0)} = \int_{-\infty}^\infty da \, e^{-4\pi {\rm{Im}}(\tau) a^2} (2a)^2 \frac{H(2ia) H(-2ia)}{(H(ia) H(-ia))^4}
~.
\eeq
The mathematical identity
\beq
\label{SU(2)db}
\log\left(\frac{H(2ia) H(-2ia)}{(H(ia) H(-ia))^4}\right) 
= -8 \sum_{k=2}^{\infty} \frac{\zeta(2k-1)}{k}(2^{2k-2}-1)(-1)^k a^{2k}
\eeq
implies the perturbative expansion (see also \cite{Russo:2012kj})
\beq
\label{SU(2)dc}
	Z_{S^4}^{(0)} = \frac{1}{4\pi(\rm{Im} \tau)^{3/2}}\left(1 -\frac{45\, \zeta(3)}{(4\pi\, \rm{Im}\tau)^2} 
	+ \frac{525\, \zeta(5)}{(4\pi\,\rm{Im}\tau)^3} + \ldots \right)
~.
\eeq
Then, employing \eqref{SU(2)bb} and the recursive \tts equations \eqref{SU(2)af} we deduce the pertubative
expansion of the 2-point functions of any chiral primary. For the first five chiral primaries the specific expressions 
are
\begin{align}
	\label{eq:g2loop2}
	g_2^{(0)} & = \frac38 \frac{1}{(\rm{Im} \tau)^2} - \frac{135 \,\zeta(3)}{32 \, \pi^2} \frac{1}{(\rm{Im} \tau)^4} 
	+ \frac{1575 \,\zeta(5)}{64 \, \pi^3} \frac{1}{(\rm{Im} \tau)^5} + \ldots~,\\
	\label{eq:g4loop2}	
	g_4^{(0)} & = \frac{15}{32} \frac{1}{(\rm{Im} \tau)^4} - \frac{945 \,\zeta(3)}{64 \, \pi^2} \frac{1}{(\rm{Im} \tau)^6} 
	+ \frac{7875 \,\zeta(5)}{64 \, \pi^3} \frac{1}{(\rm{Im} \tau)^7} + \ldots~,\\
	g_6^{(0)} & = \frac{315}{256} \frac{1}{(\rm{Im} \tau)^6} - \frac{76545 \,\zeta(3)}{1024 \, \pi^2} \frac{1}{(\rm{Im} \tau)^8}
	+ \frac{1677375 \,\zeta(5)}{2048 \, \pi^3} \frac{1}{(\rm{Im} \tau)^9} + \ldots~,\\
	g_8^{(0)} & = \frac{2835}{512} \frac{1}{(\rm{Im} \tau)^8} - \frac{280665 \,\zeta(3)}{512 \, \pi^2} 
	\frac{1}{(\rm{Im}\tau)^{10}} 
	+ \frac{1913625 \,\zeta(5)}{256 \, \pi^3} \frac{1}{(\rm{Im} \tau)^{11}} + \ldots~,\\
	\label{eq:g10loop2}
	g_{10}^{(0)} & = \frac{155925}{4096} \frac{1}{(\rm{Im} \tau)^{10}} - \frac{91216125 \,\zeta(3)}{16384 \, \pi^2} \frac{1}{(\rm{Im} \tau)^{12}} + \frac{2982065625 \,\zeta(5)}{32768 \, \pi^3} \frac{1}{(\rm{Im} \tau)^{13}} + \ldots~.
\end{align}
In section \ref{checks} we verify independently the validity of the first two orders of these expressions (for arbitrary 
$g_{2n}^{(0)}$) in perturbation theory. For each of these 2-point functions, the leading order term comes from a 
tree-level computation. The one-loop contribution is always vanishing and the next-to-leading order contribution 
comes from a two-loop computation.

\vspace{0.5cm}
The corresponding 3-point functions follow immediately from equation \eqref{SU(2)cab}. 
In the alternative conventions \eqref{SU(2)ca} they follow from a
straightforward application of equation \eqref{SU(2)cb}. The first few coefficients are 
\begin{align}
	\hat C_{2\, 2\, 4}^{(0)} & = \sqrt{\frac{10}{3}}\left(1 - \frac{9\,\zeta(3)}{2\pi^2} \frac{1}{(\rm{Im} \tau)^2} 
	+ \frac{525 \, \zeta(5)}{8 \pi^3} \frac{1}{(\rm{Im} \tau)^3} + \ldots \right)~,\\
	\hat C_{2\, 4\, 6}^{(0)} & = \sqrt{7}\left(1 - \frac{9\,\zeta(3)}{\pi^2} \frac{1}{(\rm{Im} \tau)^2} 
	+ \frac{675 \, \zeta(5)}{4 \pi^3} \frac{1}{(\rm{Im} \tau)^3} + \ldots \right)~,\\
	\hat C_{2\, 6\, 8}^{(0)} & = 2\sqrt{3}\left(1 - \frac{27\,\zeta(3)}{2\pi^2} \frac{1}{(\rm{Im} \tau)^2} 
	+ \frac{2475 \, \zeta(5)}{8 \pi^3} \frac{1}{(\rm{Im} \tau)^3} + \ldots \right)~,\\
	\hat C_{2\, 8\, 10}^{(0)} & = \sqrt{\frac{55}{3}}\left(1 - \frac{18\,\zeta(3)}{\pi^2} \frac{1}{(\rm{Im} \tau)^2} 
	+ \frac{975 \, \zeta(5)}{2 \pi^3} \frac{1}{(\rm{Im} \tau)^3} + \ldots \right)~,\\
	\hat C_{4\, 4\, 8}^{(0)} & = 3\sqrt{\frac{14}{5}}\left(1 - \frac{18\,\zeta(3)}{\pi^2} \frac{1}{(\rm{Im} \tau)^2} 
	+ \frac{825 \, \zeta(5)}{2 \pi^3} \frac{1}{(\rm{Im} \tau)^3} + \ldots \right)~,\\
	\hat C_{4\, 6\,10}^{(0)} & = \sqrt{66}\left(1 - \frac{27\,\zeta(3)}{\pi^2} \frac{1}{(\rm{Im} \tau)^2} 
	+ \frac{2925 \, \zeta(5)}{4 \pi^3} \frac{1}{(\rm{Im} \tau)^3} + \ldots \right)
~.
\end{align}

\subsubsection*{1-Instanton sector}

The contribution of instantons can be deduced from known expressions of $Z_{\rm inst}$ without much 
additional effort. For example, in the 1-instanton sector\footnote{By this we mean contributions of 1 instanton 
or 1 anti-instanton, i.e. the part that scales like $\exp\left(-{8 \pi^2 \over g_{YM}^2}\right)$.}
the first few orders in the perturbative expansion of $Z_{S^4}$ are 
\begin{equation}
\label{SU(2)ea}
Z_{S^4}^{(1)} =  \cos\theta \exp\left(-{8 \pi^2 \over g_{_{YM}}^2}\right)
\left(-{3 \over 4 \pi({\rm Im}\tau)^{3/2}}\right)
\bigg[1- {1\over 8 \pi {\rm Im}\tau} - {45 \zeta(3) \over 16 \pi^2 ({\rm Im}\tau)^2} 
+{105(\zeta(3) + 10 \zeta(5)) \over 128 \pi^3 ({\rm Im}\tau)^3} + \ldots\bigg]
.
\end{equation}
We have written out $\theta = \pi (\tau+\overline{\tau})$ and $g_{YM}$ explicitly in some of the terms, to make the 
expression more intuitive. The corresponding corrections $g_{2n}^{(1)}$ of $g_{2n}$ can be computed by starting 
with \eqref{SU(2)bb} 
\begin{align}
\label{oneinst}
 g_2 = \partial_\tau \partial_{\overline{\tau}} \log\left(\, Z_{S^4}^{(0)}+ Z_{S^4}^{(1)}+  \ldots \right)
~,
\end{align}
recursively applying \eqref{SU(2)af} 
\be
\label{SU(2)instantonaf}
	g_{2n+2} = g_{2n}\, \partial_{\tau}\partial_{\bar \tau} \log g_{2n} + \frac{g_{2n}^2}{g_{2n-2}} + g_2\, g_{2n},
	~~ n=1,2,\ldots
\ee
and finally isolating the $\exp\left(-{8 \pi^2 \over g_{_{YM}}^2}\right)$ contribution $g_{2n}^{(1)}$ at every level 
$g_{2n}$.  For the first terms we find
\begin{align}
	g_2^{(1)} & = \cos\theta \,\exp\left(-{8 \pi^2 \over g_{_{YM}}^2}\right)\left({3\over 8 ({\rm Im}\tau)^2} 
	+ {3 \over 16 \pi ({\rm Im}\tau)^3} -{135 \zeta(3)\over32 \pi^2 ({\rm Im}\tau)^4} +\ldots \right),\\
	g_4^{(1)} & = \cos\theta \exp\left(-{8 \pi^2 \over g_{_{YM}}^2}\right)\left({15 \over 16 ({\rm Im}\tau)^4} 
	+ {15 \over 32 \pi ({\rm Im}\tau)^5} -{945 \zeta(3) \over 32 \pi^2 ({\rm Im}\tau)^6} +\ldots\right)
\end{align}
It is straightforward to continue with higher $n$ if desired. Analogous results can be obtained likewise for the 
general $\ell$-instanton sector. From these 2-point functions we can also express the exact instanton corrections to chiral 
primary 3-point functions.

It would be interesting to confirm these results with an independent perturbative computation in 
the $\ell$-instanton sector. 

\subsubsection{Comments on $SL(2,{\mathbb Z})$ duality}

It is interesting to explore the transformation properties of correlators of chiral primaries in $\NN=2$ SCQCD under
non-perturbative $SL(2,{\mathbb Z})$ transformations
\be
\label{sl2z}
\tau' = \frac{a \tau + b}{c\tau + d}\qquad,\qquad a,b,c,d\in {\mathbb Z}\quad,\quad ad-bc = 1
~.
\ee
We expect that the Zamolodchikov metric obeys the identity
\begin{align}
G_{\tau'\,\overline{\tau}'}\,\,d\tau' d\overline{\tau}' = G_{\tau \,\overline{\tau}} d\tau d\overline{\tau}
~,
\end{align}
or equivalently
\be
\label{metricsl2}
G_{\tau \overline{\tau}} \left(\frac{a \tau + b}{c\tau+ d} , \frac{a \overline{\tau} + b}{c\overline{\tau}+ d}\right)   =   |c\tau + d|^2 G_{\tau \, \overline{\tau}} \,(\tau,
\overline{\tau})
~.
\ee
A similar transformation property holds for the 2-point function $g_2 = G_{\tau \overline{\tau}}/192$. 

Given the relation between the Zamolodchikov metric and the $S^4$ partition function
\be
\label{zzz}
G_{\tau\overline{\tau}} = 192 \,\,\partial_\tau \partial_{\bar \tau} Z_{S^4}
\ee
and taking into account the transformation \eqref{sl2z}, 
we notice that the validity of \eqref{metricsl2} requires the partition function $Z_{S^4}$ 
to be $SL(2,{\mathbb Z})$ invariant up to K\"ahler transformations
\beq
\label{zzza}
\log Z_{S^4}(\tau') = \log Z_{S^4}(\tau) + f(\tau) + \overline{f(\tau)}
~.
\eeq
The issue we would like to address here is the following:
suppose that we have verified the correct $SL(2,{\mathbb Z})$ transformation of $g_2$. 
What is the $SL(2,{\mathbb Z})$ behavior of the
2-point functions $g_{2n}$ of the higher order chiral primaries? 

The \tts equations provide a specific answer.
Assuming $g_2' = |c\tau+d|^2 g_2$, it is easy to verify recursively from \eqref{SU(2)af} that
\be
\label{slgn}
g_{2n}' = |c\tau+d|^{2n} g_{2n}
~.
\ee
Alternatively, in the normalization \eqref{SU(2)ca}, equations \eqref{SU(2)cb} and \eqref{slgn} 
imply that the 3-point functions are $SL(2,{\mathbb Z})$ invariant
\beq
\label{slgna}
\hat C_{2m\, 2n \, \overline{2m+2n}}' =\hat C_{2m\, 2n \, \overline{2m+2n}}~,
\eeq
which is consistent with expectations. See \cite{Gomis:2009xg} for a related discussion of the $S$-duality 
properties of chiral primary correlation functions in $\NN=4$ SYM theory.

\subsection{$SU(N)$ with $2N$ hypermultiplets}
\label{SU(N)}

The case of general $SU(N)$ gauge group can be analyzed in a similar fashion. 
Unfortunately, for general $N\geq 3$ it is less clear under which conditions we can identify the relevant solution of the \tts equations. 
We proceed to discuss the detailed structure of the $SU(N)$ \tts equations.

The general $SU(N)$ $\NN=2$ SCQCD theories possess $N-1$ chiral ring generators represented by 
the single-trace operators
\beq
\label{SU(N)aa}
\Tr [\varphi^2]~, ~~ \Tr[\varphi^3]~,~~ \cdots ~, ~~ \Tr[\varphi^N]
~.
\eeq
The general element of the chiral ring is freely generated from these operators and
can be viewed as a linear combination of the primaries
\beq
\label{SU(N)ab}
\phi_{(n_1, n_2,\ldots, n_{N-1})} \propto 
\prod_{i=1}^{N-1} \left( \Tr [\varphi^{i+1}] \right)^{n_i}
~.
\eeq
The operator that gives rise to the single exactly marginal direction $\OO_\tau$ of the theory is
\beq
\label{SU(N)ac}
\phi_2 \equiv \phi_{(1,0,\ldots,0)}
~.
\eeq
We notice that the scaling dimension of the generic chiral primary \eqref{SU(N)ab} is 
$\Delta = \sum_{i=1}^{N-1} (i+1)n_i$. Obviously, there are values of $\Delta$ where more
than one chiral primary can have the same scaling dimension. Such chiral primaries can 
mix non-trivially with each other to exhibit non-diagonal $\tau$-dependent 2-point 
function matrices. We verify this mixing explicitly in specific examples at tree-level in 
subsection \ref{3tree}. 

The OPE of the chiral primaries \eqref{SU(N)ab} can be chosen to take the form
\beq
\label{SU(N)ad}
\phi_{(n_1,\ldots,n_{N-1})}(x) \, 
\phi_{(m_1,\ldots,m_{N-1})}(0) 
= \phi_{(n_1+m_1,\ldots,n_{N-1}+m_{N-1})}(0)+\ldots
~,
\eeq 
or in more compact notation
\beq
\label{SU(N)ae}
\phi_K(x)\, \phi_L(0) = \phi_{K+L}(0) +\ldots
~.
\eeq
This choice allows us to fix the non-vanishing OPE coefficients to
\beq
\label{SU(N)aea}
C_{K\, L}^{K+L} =1~,
\eeq
in analogy to the $SU(2)$ equation \eqref{SU(2)ac}. In this way, once we choose the normalization of 
the chiral ring generators \eqref{SU(N)aa} the normalization of all the chiral primary fields
is uniquely determined. We will consider a normalization of $\phi_2$ that adheres to the conventions 
\eqref{actiondeformation}, \eqref{f192}. The remaining chiral primaries in \eqref{SU(N)aa}
are chosen with an arbitrary normalizing factor $\NN_K(\tau)$ that is a holomorphic function
of the complex coupling $\tau$.

\subsubsection{The structure of the $SU(N)$ \tts equations}

In these conventions the \tts equations \eqref{maintt} become
\beq
\label{SU(N)af}
\partial_{\bar \tau} \left( g^{\bar M_\Delta L_\Delta} \partial_\tau g_{K_\Delta \bar M_\Delta} \right)
=g_{K_\Delta +2, \bar R_\Delta +\bar 2}\, g^{\bar R_\Delta L_{\Delta}}
-g_{K_\Delta \bar R_\Delta}\, g^{\bar R_\Delta-\bar 2, L_\Delta-2}
-g_2 \, \delta^{L_\Delta}_{K_\Delta}
~.
\eeq
The addition of 2 in the index notation $K+ 2$ refers to the element $\phi_2 \, \phi_K$.
The subindex $\Delta$ on the indices has been added here to flesh out the scaling dimension of the 
corresponding chiral primaries. Sample tree-level
checks of equations \eqref{SU(N)af} (that exhibit the non-trivial mixing of chiral primaries) 
are collected in section \ref{3tree}.

Similar to the $SU(2)$ case the equations \eqref{SU(N)af} relate 2-point functions of
chiral primaries at three different scaling dimensions and can be recast in the recursive form
\beq
\label{SU(N)afa}
g_{K_\Delta+2, \bar N_\Delta + \bar 2}
= g_{L_\Delta \bar N_\Delta} 
\partial_{\bar\tau} \left( g^{\bar M_\Delta L_\Delta} \partial_\tau g_{K_\Delta \bar M_\Delta} \right)
+g_{K_\Delta \bar M_\Delta} g^{\bar M_\Delta -\bar 2, L_\Delta -2} g_{L_\Delta \bar N_\Delta}
+g_2 \, g_{K_\Delta \bar N_\Delta}
~.
\eeq
However, unlike the situation of the $SU(2)$ gauge group, the complicated degeneracy pattern of the 
general $SU(N)$ theory and the corresponding non-trivial mixing of the chiral primary fields makes 
this system of differential equations a far more complicated one
to solve explicitly in terms of a few externally determined data (like the Zamolodchikov metric). 

Most notably, the LHS of equation \eqref{SU(N)afa}
involves primaries that belong in a subsequence generated by multiplication with the field $\phi_2$.
In contrast, the RHS involves in general 2-point functions of all available chiral primaries. 
This feature complicates the recursive solution of the system of equations \eqref{SU(N)afa}. 
As we move up in scaling dimension with the action of $\phi_2$ the number of degenerate fields 
will stay the same or increase. Increases are due to the appearance of additional degenerate chiral
primary fields that involve the action of the extra chiral ring generators other than $\phi_2$, i.e. $\Tr[\varphi^3]$
$etc$. In such cases, there are seemingly new 2-point function coefficients that have not been determined 
recursively from the previous lower levels and represent new data that need to be provided externally. 
It is an interesting open question whether other properties (like the property of positivity over 
the entire moduli space) are strong enough to reduce the number of unknowns and fix the full solution
uniquely.

Despite the apparent complexity of \eqref{SU(N)afa}, it is quite likely that this system has a hidden 
structure that allows to simplify its description. For example, in section \ref{Ntree} we find 
preliminary evidence at tree-level that one can isolate differential equations that form a closed system on 
the subsequence of the chiral primary fields $(\phi_2)^n$. If true, the data of such subsequences could
be determined solely in terms of the $SU(N)$ $S^4$ partition function in direct analogy to the $SU(2)$ case.
Such possibilities are currently under investigation.

\subsubsection{3-point functions}

The non-vanishing 3-point structure constants of the $SU(N)$ theory are
\beq
\label{SU(N)caa}
C_{K_{\Delta_1}\, L_{\Delta_2} \, \overline{ M}_{\Delta_1+\Delta_2}} = 
C_{K_{\Delta_1}\, L_{\Delta_2}}^{(K+L)_{\Delta_1+\Delta_2}} 
g_{(K+L)_{\Delta_1+\Delta_2},\overline{M}_{\Delta_1+\Delta_2}}
=g_{(K+L)_{\Delta_1+\Delta_2},\overline{M}_{\Delta_1+\Delta_2}}
~.
\eeq
This relation is the $SU(N)$ generalization of \eqref{SU(2)caa}, \eqref{SU(2)cab}.
Consequently, a solution of the \tts equations \eqref{SU(N)af} determines immediately also 
the 3-point functions \eqref{SU(N)caa}.

The conversion of the above results into the language of the common alternative normalization 
\eqref{SU(2)ca}
\beq
\label{SU(N)ba}
\left< \hat\phi_K(x) \overline{\hat \phi}_{\bar L}(0) \right>
= \frac{\delta_{K\bar L}}{|x|^{2\Delta_{K}}}
\eeq
requires a transformation
\beq
\label{SU(N)bb}
\hat \phi_K = \sum_L \NN_K^{~~L} \phi_L
\eeq
at each scaling dimension $\Delta$, where the matrix elements $\NN_K^{~~L}$ are 
suitable functions of the 2-point coefficients $g_{K\bar L}$.
Once the matrix elements $\NN_K^{~~L}$ are determined the 3-point structure constants
$\hat C_{IJ\bar K}$ in the basis \eqref{SU(N)ba} can be written as
\beq
\label{SU(N)bc}
\hat C_{IJ\bar K} = \sum_{L_1, L_2, \bar L_3} \NN_I^{~~L_1} \NN_J^{~~L_2} \bar \NN_{\bar K}^{~~\bar L_3}
g_{L_1+L_2, \bar L_3}
~.
\eeq

\section{Checks in perturbation theory}
\label{checks}

In this section we perform a number of independent checks of the above statements in 
perturbation theory. These checks provide a concrete verfication 
of the validity of the general formal proof of the \tts equations in \cite{Papadodimas:2009eu},
and allow us to verify that the \tts equations were applied  correctly in the previous section.
In the process, we encounter and comment on several individual properties of correlation
functions in $\NN=2$ SCQCD. We work in the conventions listed in appendix \ref{conventions}.

\subsection{$SU(2)$ SCQCD}
\label{SU(2)pert}

We begin with a perturbative computation up to 2 loops of the 2-point coefficients $g_{2n}$ in the $SU(2)$
$\NN=2$ SCQCD theory.

\subsubsection{Tree-level}
\label{2tree} 

Let us start with a comment about normalizations in the general $SU(N)$ theory.
At leading order in the weak coupling limit, $g_{YM} \ll 1$, (and the conventions summarized in 
appendix \ref{conventions}) the 2-point function of the adjoint scalars $\varphi = \varphi^a T^a$ is
\beq
\label{2treeaa}
\left< \varphi^a (x) \overline{\varphi}^b(0) \right> = \delta^{ab} \frac{1}{\pi \, \rm Im\tau} \frac{1}{|x|^2}
~.
\eeq
$T^a$ ($a=1,\ldots, N^2-1$) is a basis of the $SU(N)$ Lie algebra.
Normalizing the chiral primary operator $\phi_2$ as
\beq
\label{2treeab}
\phi_2 = \frac{\pi}{4N} \Tr [\varphi^2] = \frac{\pi}{4} \varphi^a \varphi^a
\eeq
we obtain 
\beq
\label{2treeac}
\left< \phi_2 (x) \overline{\phi}_2(0) \right> = \frac{N^2-1}{8} \frac{1}{({\rm Im} \tau)^2} \frac{1}{|x|^4}
~.
\eeq
On the other hand, the exactly marginal operator $\OO_\tau$, \eqref{sqcdaba}, has the explicit form 
presented in equation \eqref{Aae} of appendix \ref{conventions}. A tree-level computation yields\footnote{At 
tree-level only the gauge part $\frac{i\pi}{16}  F^a_{\mu\nu+} F^{\mu\nu+a} $ of $\OO_\tau$
in \eqref{Aae} contributes. The auxiliary fields contribute only contact terms and the cubic interactions are 
subleading in $g_{YM}$. The boson and fermion kinetic terms vanish on-shell. A similar observation was made
in \cite{Intriligator:1998ig}.}
\beq
\label{2treead}
\left< \OO_\tau(x) \overline{\OO}_{\tau}(0) \right> = 24 (N^2-1) \frac{1}{({\rm Im}\tau)^2} \frac{1}{|x|^8}~,
\eeq
which is consistent with the conventions \eqref{zdef}, \eqref{ctwopoint}, \eqref{f192}.
This is important for the validity of the \tts equations \eqref{maintt}, 
or the equations \eqref{SU(2)ae} in the $SU(2)$ case of this subsection. 

Specializing now to the $SU(2)$ case we find that the 2-point function \eqref{2treeac} has the tree-level
coefficient
\beq
\label{2treeae}
g_2 = \frac{3}{8} \frac{1}{({\rm Im }\tau)^2}
~.
\eeq
We can read off the 2-point function coefficients $g_{2n}$ of the higher chiral primary operators 
$\phi_{2n}= (\phi_2)^n$ from free field Wick contractions in the 2-point correlation function
\beq
\label{2treeaf}
\left< \phi_{2n}(x) \, {\overline\phi}_{2n}(0) \right> = \left< (\phi_2)^n(x) \, ({\overline \phi}_2)^n(0) \right>
~.
\eeq
A brute-force computation gives
\beq
\label{2treeag}
g_{2n} = \frac{(2n+1)!}{6^n} g_2^n
~.
\eeq
With this result the  \tts equations \eqref{SU(2)ae} 
\beq
\label{2treeai}
\partial_\tau \partial_{\bar\tau} \log g_{2n} 
= \frac{g_{2n+2}}{g_{2n}} - \frac{g_{2n}}{g_{2n-2}} - g_2
\eeq
reduce at tree-level to the differential equation
\beq
\label{2treeaj}
\partial_\tau \partial_{\bar \tau} \log g_2 = \frac{4}{3} g_2~,
\eeq
which is found to hold for the $g_2$ given in equation \eqref{2treeae}.

\subsubsection{Quantum corrections up to 2 loops}
\label{sec:loop}

We proceed to compute the first non-vanishing quantum corrections to $g_{2n}$ in perturbation theory. This will 
allow us to reproduce the Zamolodchikov metric derived from localization \cite{Gerchkovitz:2014gta} at $g_{YM}^4$ 
order and will provide a test of the \tts equations at the quantum level. Furthermore, due to the discussion in section 
\ref{sec:2and3ptf}, this provides a $g_{YM}^4$ check of the chiral primary three-point functions in a diagonal basis 
as well. We will use the techniques of \cite{Andree:2010na}, namely we will exploit the fact that quantum corrections 
for $\mathcal{N}=4$ SYM vanish at each order in perturbation theory\footnote{See 
\cite{Penati:1999ba, Penati:2000zv} for perturbative computations of 2-point functions of chiral primaries in 
${\cal N}=4$ SYM.}, so that we only need to compute the diagrammatic difference between the 
$\mathcal{N}=2$ and $\mathcal{N}=4$ theories.

Following \cite{Andree:2010na}, it is easy to see that the diagrammatic difference between $\mathcal{N}=2$ and 
$\mathcal{N}=4$ at order $g_{YM}^2$ vanishes. It immediately follows that the theory does not receive quantum 
corrections to this order, consistent with the results from localization \eqref{eq:g2loop2}-\eqref{eq:g10loop2}.
 
We now examine the diagrams that contribute to order $g_{YM}^4$ to the 2-point function
\beq
\label{2treeafbb}
\left< \phi_{2n}(x) \, {\overline\phi}_{2n}(0) \right> = \left< (\phi_2)^n(x) \, ({\overline \phi}_2)^n(0) \right> 
= {g_{2n} \over |x|^{4n}}
~.
\eeq
To understand what type of diagrams can contribute to this order, it is convenient to temporarily regard the adjoint 
scalar $\varphi$ lines as external and change the normalization of the fields so that the coupling constant 
dependence is on the vertices. Diagrams which differ between $\mathcal{N}=2$ and 
$\mathcal{N}=4$ must involve hypermultiplets running in the internal lines. After a brief inspection of 
the ${\cal N}=2$ SCQCD Lagrangian it is not too hard to convince oneself that the only possible types of diagrams that can contribute to order $g_{YM}^4$ (and which differ between ${\cal N}=2$ and ${\cal N}=4$) come from two types of topologies, when trying to connect the $2n$ `external lines' of $\varphi$ at point $x$ to the $2n$ 
`external lines' of $\overline{\varphi}$ at point $0$:
\begin{itemize}
\item[$a)$] diagrams where one external $\varphi$ line is connected to one external $\overline{\varphi}$ line
by a 2-loop-corrected $\varphi-\overline{\varphi}$ propagator, while all others lines are connected by 
free propagators
\item[$b)$] diagrams where two external $\varphi$ lines and two external $\overline{\varphi}$ lines are all connected together by a nontrivial 4-leg subdiagram, while the remaining $\varphi$ and $\overline{\varphi}$ lines are connected by free propagators. 
\end{itemize}

\noindent Let us examine the former first. We denote the quantum corrected propagator as
\begin{align}
	\label{eq:defa}
	\left<\varphi^a(x) \bar\varphi^b(y)\right> = \delta^{ab}S(x-y) 
	= \delta^{ab}S^{(0)}(x-y) \big(1 + f_1 \,g_{YM}^4 + \ldots\big)~,
\end{align}
where $S^{(0)}(x-y)$ is the tree-level propagator \eqref{2treeaa} and we have used the fact that the $g_{YM}^4$ 
corrections are proportional to the tree-level propagator \cite{Andree:2010na}. $f_1$ is a numerical constant that 
we will determine in the following.
\begin{figure}
	\begin{center}
		\includegraphics[scale=0.45]{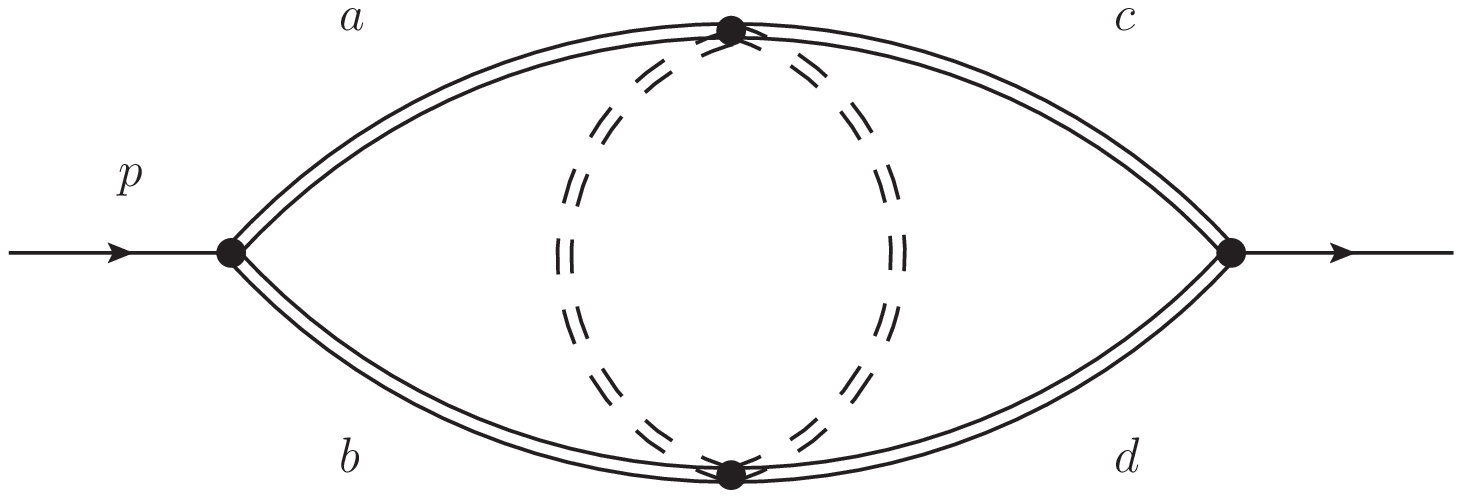}
		\hspace{1cm}
		\includegraphics[scale=0.45]{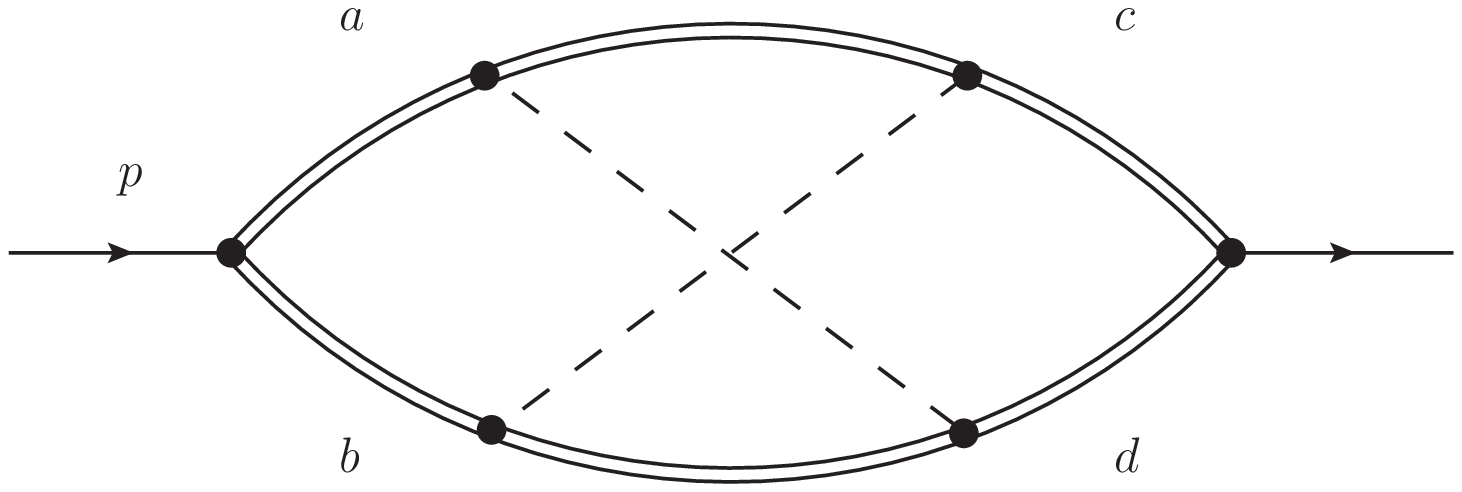}
	\end{center}
	\begin{center}
		$D_1$
		\hspace{7.5cm}
		$D_2$
	\end{center}
	\caption{The diagrams $D_1$ and $D_2$. Solid double lines represent $\varphi$ propagators, dashed 
	double lines correspond to hyperscalars and dashed lines to hyperfermions. $a,b,c$ and $d$ are adjoint
	gauge indices and $p$ is the incoming momentum. 
	}
	\label{fig:2diag}
\end{figure}

Regarding diagrams of type $b)$, there are only two  diagrams\footnote{We remind the reader that we are only considering diagrams which 
differ between ${\cal N}=2$ SCQCD and ${\cal N}=4$ SYM with the same gauge group. Also the statement that 
these are the only diagrams is true only for $SU(2)$ gauge group. See \cite{Andree:2010na} for useful 
background.} that can contribute to this order, which 
are shown in figure \ref{fig:2diag}. In the diagram $D_1$ hyperscalars run in the internal loop, while the diagram $D_2$ corresponds 
to the exchange of hyperfermions. In more detail, we define $D_1(x,y)$ and $D_2(x,y)$ as
\begin{align}
	D_1(x,y) & 
	= \frac12 \left<\varphi^a(x)\varphi^b(x)\bar\varphi^c(y)\bar\varphi^d(y) (\Xi_{1})^2\right>_{\rm connected}~,\\
	D_2(x,y) & 
	= \frac{1}{4!} \left<\varphi^a(x)\varphi^b(x)\bar\varphi^c(y)\bar\varphi^d(y) (\Xi_{2})^4\right>_{\rm connected}~,
\end{align}
where $\Xi_{1}$ and $\Xi_{2}$ are the interaction actions associated to the terms in the Lagrangian \eqref{Aacc}  
coupling the vector sector to the hypermultiplet sector, namely
\begin{align}
	\Xi_{1} & = \int d^4 x\, Q_\II \left(\bar\varphi \varphi +\varphi\bar\varphi \right) Q^\II~,\\
	\Xi_{2} & = i\sqrt{2}\int d^4 x\,(\tilde\psi \varphi \psi - \bar\psi \bar\varphi \bar{\tilde\psi})~,
\end{align}
and we take Wick's contractions that correspond to connected diagrams only.

It is easy to see that all the other diagrams either vanish or are identical to their $\mathcal{N}=4$ counterparts.
We start by examining the gauge structure of these diagrams. Both are proportional to 
$\mathrm{Tr} (T^a T^c T^b T^d)$ (or permutations thereof), so the difference between the $\mathcal{N}=2$ and 
$\mathcal{N}=4$ color factors reads
\begin{align}
	4\, \mathrm{Tr} (T^a T^c T^b T^d)_{\rm{fundamental}} - \mathrm{Tr} (T^a T^c T^b T^d)_{\rm{adjoint}} 
	= -\frac12 \left(\delta^{ac}\delta^{bd} + \delta^{ad}\delta^{bc} + \delta^{ab}\delta^{cd}\right)~,
\end{align}
where the factor of 4 in the equation above comes from the fact that the $\mathcal{N}=2$ theory has 4 
hypermultiplets. It is thus convenient to define the quantity
\begin{align}
\label{constantcdef}
	\mathcal{C} \equiv \delta^{ac}\delta^{bd} + \delta^{ad}\delta^{bc} + \delta^{ab}\delta^{cd}~,
\end{align}
and parametrize the contribution from these two diagrams as
\begin{align}
	\label{eq:defc}
	D_1(x,y) + D_2(x,y) = \mathcal{C}\,S^{(0)}(x-y)^2\,f_2\,g_{YM}^4~,
\end{align}
where $f_2$ is a numerical constant that we will determine momentarily.

With these results, it is straightforward to work out the combinatorics and find the $g_{YM}^4$ corrections to the 
correlation functions $g_{2n}$ as a function of the two contributions $f_1$ and $f_2$. After some work we find that 
the result is
\begin{align}
	\label{eq:g2n1loopcomb}
	\langle  \phi_{2n}(x) \,\overline{\phi}_{2n}(y)\rangle = \left(\frac{\pi}{4}\right)^{2n}(2n+1)!\,S^{(0)}(x-y)^{2n}  \Big[1 + \frac{n}{2} \left(4 f_1 + (6n - 1) f_2\right)\,g_{YM}^4\Big]~,
\end{align}
where $f_1$ and $f_2$ are defined in equations \eqref{eq:defa} and \eqref{eq:defc} respectively. In order to derive 
the expression above, one has to consider all the possible ways to connect the propagators associated to 
$\phi_{2n}$ with those associated to $\bar\phi_{2n}$, with the insertion of $g_{YM}^4$ corrections coming from 
the diagrams described above. We notice that the contribution coming from $g_{YM}^4$ diagrams with two 
external $\varphi$ lines has a different dependence on $n$ compared to the one coming from diagrams with four 
external $\varphi$ lines, reflecting the different combinatorial properties of these graphs.

It is important to notice that the equation above is \emph{not} automatically consistent with the \tts equations. In fact, 
we find that demanding that \eqref{eq:g2n1loopcomb} satisfies the \tts equations leads to the non-trivial condition
\begin{align}
\label{eq:cafromtts}
	f_2 = \frac25 f_1~.
\end{align}
We conclude that the \tts equations do encode non-trivial information about the quantum corrections to chiral 
primary correlation functions, as they are sensitive to the ratio $f_2/f_1$. Determining this ratio by explicitly 
computing the relevant Feynman diagrams will thus provide us with a stringent test of these equations at the 
quantum level.

We will now determine the value of $f_1$ and $f_2$ by computing the Feynman diagrams $D_1$ and $D_2$. 
We will show that their ratio is precisely the one predicted by the \tts equations. Furthermore, the result will allow 
us to compute the $g_{YM}^4$ correction to the Zamolodchikov metric, providing thus a perturbative check of the 
results of \cite{Gerchkovitz:2014gta}.
\subsubsection*{Computation of $f_1$ and $f_2$}
Recall that the tree-level propagator \eqref{2treeaa} reads
\begin{align}
	S^{(0)}(x-y) = \frac{g_{YM}^2}{4\pi^2 (x - y)^2} = g_{YM}^2\int \frac{d^4 p}{(2\pi)^4} \frac{e^{-i p (x-y)}}{p^2}~.
\end{align}
As is customary, we work in momentum space and in dimensional regularization, where the spacetime dimension 
$d$ is $d = 4 - 2\epsilon$.

\begin{figure}
\begin{center}
\includegraphics[scale=0.55]{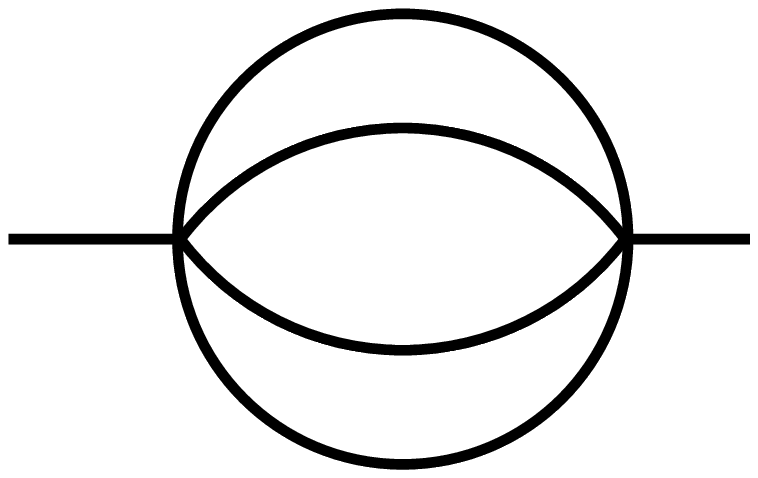}\hspace{1cm}
\includegraphics[scale=0.55]{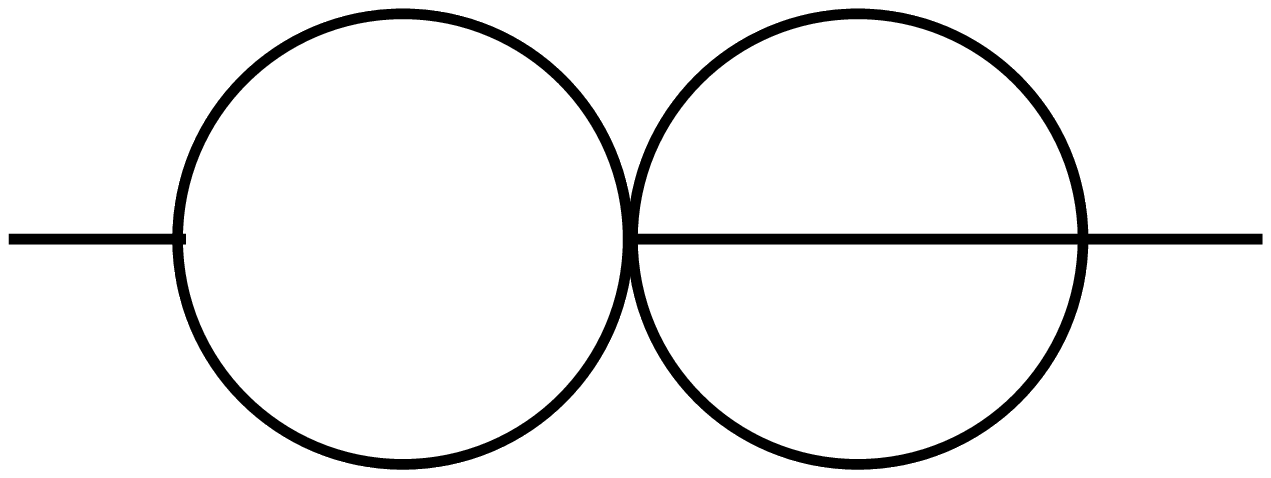}\\[0.3cm]
$B_{41}$\hspace{6cm}$B_{51}$\phantom{random}\\[1cm]
\includegraphics[scale=0.55]{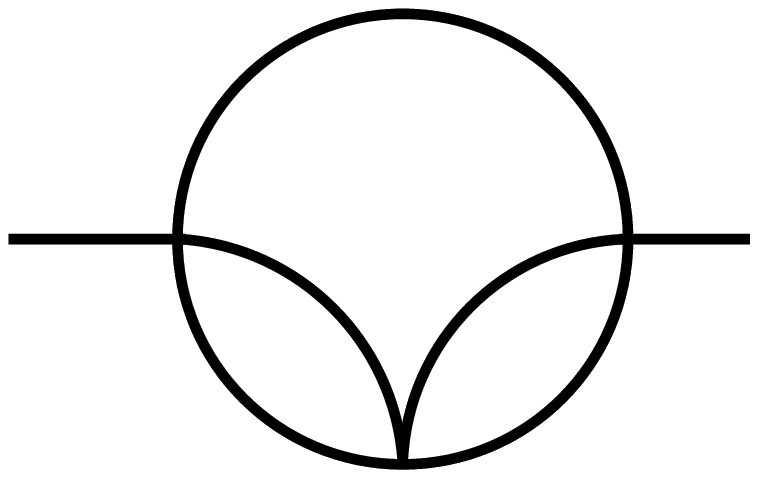}\hspace{0.5cm}
\includegraphics[scale=0.55]{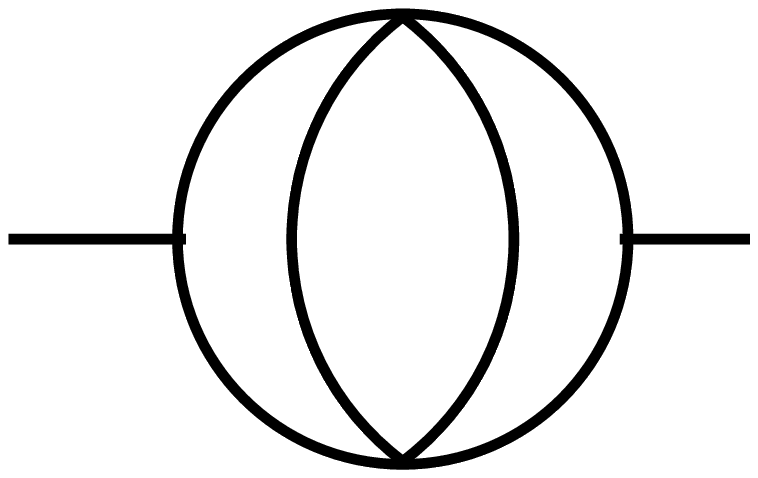}\hspace{0.5cm}
\includegraphics[scale=0.55]{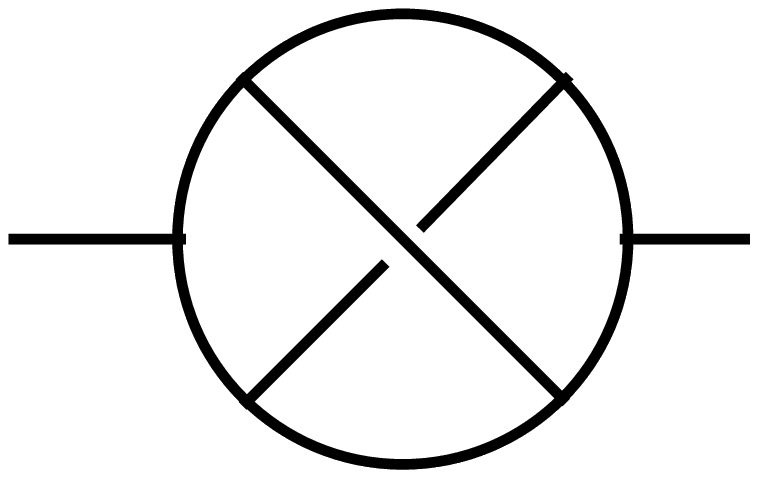}\\[0.3cm]
$B_{52}$\hspace{4.5cm}$B_{62}$\hspace{4.5cm}$B_{81}$
\end{center}
\caption{Master integrals appearing in the reduction of the Feynman diagrams of figure \ref{fig:2diag}. They 
are associated to loop integrals consisting of scalar propagators only (i.e. without non-trivial numerators).}
\label{fig:masterint}
\end{figure}

The $g_{YM}^4$ correction to the propagator $S^{(1)}(x-y)$ was computed in \cite{Andree:2010na}, and is given by
\begin{align}
	\label{eq:AYpropcorr}
	S^{(1)}(x-y) = -\frac{15 \zeta(3)}{64 \pi^4} g_{YM}^4 S^{(0)}(x-y)~,
\end{align}
which in turn implies that
\begin{align}
\label{compf1}
	f_1 =  -\frac{15 \zeta(3)}{64 \pi^4}~.
\end{align}
To compute the remaining two diagrams, we employ standard techniques \cite{Chetyrkin:1981qh} to reduce any 
3-loop loop integral to a linear combination of ``master integrals", whose $\epsilon$-expansion can be found in the 
literature. We will see in a moment that the only master integrals that we need are those that correspond to the 
topologies shown in figure \ref{fig:masterint}. For the convenience of the reader, we report here their 
$\epsilon$-expansion up to the order needed for our computation. We use the conventions of 
\cite{Gehrmann:2010ue}
\begin{align}
	\label{eq:B41eps}
	B_{41} & = p^{4-6 \epsilon}\frac{(4\pi)^{3\epsilon-6}}{\Gamma(1-\epsilon)^3}\Big(\frac{1}{36 \epsilon } 
	+ \frac{71}{216} + \frac{3115}{1296}\epsilon- \Big(\frac{7 \zeta (3)}{9}-\frac{109403}{7776}\Big) \epsilon ^2 
	+ \ldots \Big)~,\\
	\label{eq:B51eps}
	B_{51} & = p^{2-6 \epsilon}\frac{(4\pi)^{3\epsilon-6}}{\Gamma(1-\epsilon)^3}\Big(-\frac{1}{4 \epsilon^2 } 
	- \frac{17}{8\epsilon} - \frac{183}{16} +  \Big(3 \zeta(3)-\frac{1597}{32}\Big) \epsilon + \ldots \Big)~,\\
	\label{eq:B52eps}
	B_{52} & =p^{2-6 \epsilon}\frac{(4\pi)^{3\epsilon-6}}{\Gamma(1-\epsilon)^3}\Big(-\frac{1}{3 \epsilon^2 } 
	- \frac{10}{3\epsilon} - \frac{64}{3} + \Big(\frac{22 \zeta (3)}{3}-112\Big) \epsilon + \ldots \Big)~,\\
	\label{eq:B62eps}
	B_{62} & = p^{-6 \epsilon}\frac{(4\pi)^{3\epsilon-6}}{\Gamma(1-\epsilon)^3}\Big(\frac{1}{3 \epsilon^3 } 
	+ \frac{7}{3 \epsilon^2} + \frac{31}{3\epsilon} + \Big(\frac{8 \zeta (3)}{3}+\frac{103}{3}\Big) + \ldots \Big)~,\\
	\label{eq:B81eps}
	B_{81} & = p^{-4-6 \epsilon}\frac{(4\pi)^{3\epsilon-6}}{\Gamma(1-\epsilon)^3}\Big( 20 \zeta(5) + \ldots\Big)~.
\end{align}

The contributions coming from the diagrams $D_1$ and $D_2$  in momentum space will be denoted by 
$\tilde{D}_1(p)$ and $\tilde{D}_2(p)$ respectively. At the end of the computation, we transform back to position 
space using the formula
\begin{align}
\label{eq:fouriereps}
	\int \frac{d^{d}p}{(2\pi)^d} \frac{e^{-i p x}}{(p^2)^{2-k+\alpha \epsilon}} 
	= \frac{2^{2k-4}}{\pi^2}(-1)^k(k-1)!(k-2)! \alpha \frac{\epsilon}{(x^2)^{k-(\alpha+1)\epsilon}}
	\Big(1+ O(\epsilon)\Big)
~.
\end{align}
This formula tells us that we only need to determine the $1/\epsilon$ term in the Feynman diagrams of interest, 
since they are the only ones that can contribute to the finite part of the position space correlator (see also 
\cite{Penati:1999ba, Penati:2000zv}). We will explicitly show that all the higher-order poles cancel exactly 
between the two diagrams $D_1$ and $D_2$, as expected from extended supersymmetry.

We first examine the diagram $D_1$. We find that its contribution in momentum space is given by
\begin{align}
	\tilde{D}_1(p) = - 8\, g_{YM}^8\, \mathcal{C}\, B_{62}~,
\end{align}
where $B_{62}$ is the master integral associated to the topology of the corresponding diagram in figure 
\ref{fig:masterint} and $\mathcal{C}$ was defined in \eqref{constantcdef}. Since the diagram is already in the 
``master integral" form, we do not need to further reduce it and we can directly use the result in equation 
\eqref{eq:B62eps}.

The Feynman diagram $D_2$ is more complicated, but can also be reduced to a linear combination of master 
integrals as explained above. We used the {\sc mathematica} package FIRE \cite{Smirnov:2008iw} to perform the 
reduction. The result turns out to be
\begin{align}
	\tilde{D}_2(p) = \, 2\, g_{YM}^8\, \mathcal{C}\,
	\Big(& \frac{4 (2 d-5) (3 d-8) (43 d^2 - 288 d + 480)}{(d-4)^3 (2 d-7) p^4} B_{41} \nonumber\\
	&  + \frac{14  (d-3) (3 d-10) (3 d-8)}{(d-4)^2 (2 d-7) p^2} B_{51} -\frac{96 (d-3)^2}{(d-4)^2 p^2}B_{52}
	\nonumber\\
	& - \frac{(7 d^2 - 35 d + 38)}{(d-4) (2 d-7)}B_{62}+\frac{(d-4) p^4}{14-4 d}B_{81}\Big)
~.
\end{align}
Combining the results in equations \eqref{eq:B41eps}-\eqref{eq:B81eps}, we obtain
\begin{align}
	\tilde{D}_1(p) + \tilde{D}_2(p) = \Big(- \frac{8 \zeta(3)}{(4\pi)^6 \epsilon}\, g_{YM}^8\, \mathcal{C}\, 
	+ \ldots\Big) \frac{1}{p^{6\epsilon}}~,
\end{align}
where the ellipses denote terms of order $\epsilon^0$ or higher.
It is pleasing to see that the $1/\epsilon^3$ and $1/\epsilon^2$ poles precisely cancel, as well as all the 
non-$\zeta(3)$ contributions to the simple pole. Finally, we use equation \eqref{eq:fouriereps} to transform back 
to position space, so our final result reads
\begin{align}
	D_1(x,y) + D_2(x,y) = - \frac{6\zeta(3)}{64\pi^4}\, g_{YM}^4\, \mathcal{C}\, S^{(0)}(x-y)^2~.
\end{align}
Comparing with \eqref{eq:defc}, we immediately get
\begin{align}
\label{compf2}
	f_2 =  -\frac{6 \zeta(3)}{64 \pi^4}~.
\end{align}
Using the results \eqref{compf1} and \eqref{compf2} we can confirm the relation \eqref{eq:cafromtts}, which --- as 
was explained around equation \eqref{eq:g2n1loopcomb} --- implies the validity of the \tts equations for the entire
chiral ring 2-point functions $g_{2n}$ up to the relevant order!

Moreover, using equation \eqref{eq:g2n1loopcomb} we are able to provide an independent derivation of the 
$g_{YM}^4$ perturbative correction to the Zamolodchikov metric
\begin{align}
	\left<\phi_2(x)\overline{\phi}_2(y)\right> = \frac{3\pi^2}{8} \,S^{(0)}(x-y)^2 
	\Big(1 - \frac{45 \,\zeta(3)}{4 \, \pi^2} \frac{1}{(\rm{Im} \tau)^2} + \ldots\Big)~.
\end{align}
Recalling that the tree-level propagator is given by equation \eqref{2treeaa}, we find perfect agreement with the 
result from localization \eqref{eq:g2loop2} and the prediction of \cite{Gerchkovitz:2014gta}.

\subsection{$SU(N)$ SCQCD at tree level}
\label{pertchecks}

We continue with a tree-level investigation of the \tts equations for the general $SU(N)$ group.
The 2- and 3-point functions entering in \eqref{SU(N)af} can be computed directly by straightforward
Wick contractions. Examples of such computations will be provided below. 

However, before we enter these examples it is worth making first the following general point. 
Although the explicit implementation of Wick contractions can be rather cumbersome with complicated 
combinatorics, it is trivial to obtain the $\tau$-dependence of the 2-point function at leading order in the 
weak coupling limit. In general,
\be
\label{treeSub}
g_{K\overline{M}}\Big|_{\rm tree}  = {1\over ({\rm Im \tau})^{\Delta_K}} \widetilde{g}_{K\overline{M}} 
\ee
where $\widetilde{g}_{K\overline{M}}$ is coupling constant independent and contains the combinatorics from the 
contractions of the traces. From this expression the LHS of the \tts equations \eqref{SU(N)af} follows trivially as
\be
\label{LHStree}
\partial_{\overline{\tau}}\left( g^{\overline{M}L} \partial_{\tau} g_{K\overline{M}}\right)\Big|_{\rm tree} 
=  - { \Delta_K \over (\tau-\bar\tau)^2}\delta_K^L = {R\over 8  ({\rm Im}\tau)^2}\delta_K^L
\ee
where we set $\Delta_K= R/2$.

The RHS of the \tts equations \eqref{SU(N)af} has the form 
\be
\label{RHS}
[C_2,\overline{C}_{\overline{2}}]_K^L - g_2 \delta_K^L
~.
\ee
Notice that the {\it tree level} 2- and 3-point functions in this expression are exactly the same
as the ones we encountered in section \eqref{N4} in the context of $\NN=4$ SYM theory. 
As a result, we can use the identity 
\eqref{grouplemma} to recast \eqref{RHS} into the simpler form
\be
\label{RHStree}
[C_2,\overline{C}_{\overline{2}}]_K^L - g_2\, \delta_K^L\big|_{\rm tree} 
= \delta_K^L {R \over {\rm dim}\,{\cal G}} g_2\big|_{\rm tree} 
= {R \over N^2-1} {N^2-1 \over 8} {1 \over ({\rm Im}\tau)^2} 
\delta_K^L =  {R\over 8 ({\rm Im}\tau)^2} \delta_K^L
~.
\ee
We used the fact that for the $SU(N)$ theories ${\rm dim}\,{\cal G} = N^2-1$. 
Comparing the LHS \eqref{LHStree} and the RHS \eqref{RHStree} we find that the \tts equations 
are obeyed at tree level for any $SU(N)$ ${\cal N}=2$ SCFT and for all sectors of charge $R$ in the chiral ring.

The reader should appreciate that the short argument we have just presented is simpler than the
general proof of the \tts equations in \cite{Papadodimas:2009eu} because it makes explicit use of the special 
properties of correlators in a free CFT, such as the tree-level identity \eqref{grouplemma}, and its proof in 
appendix \ref{LemmaApp}.

\subsection*{$SU(3)$ examples}
\label{3tree}

To illustrate the content of the above equations and the new features of the $SU(N)$ \tts 
equations $(N\geq 3)$ (compared to the $SU(2)$ case) we consider a few sample tree-level
computations in the $SU(3)$ theory.

The $SU(3)$ $\NN=2$ SCQCD theory possesses two chiral ring generators, 
$\phi_2$ and $\phi_3$. We normalize $\phi_2$ as in \eqref{2treeac} and $\phi_3$ as 
\beq
\label{Caa}
\phi_3 = \frac{\NN_3}{8} \, \Tr [ \varphi^3 ]
\eeq
with an arbitrary $\tau$-independent normalization constant $\NN_3$.

The \tts equation \eqref{SU(N)afa} applied to scaling dimension $\Delta=3$ gives
\beq
\label{Cab}
g_5 = g_3\, \partial_\tau \partial_{\bar\tau} \log g_3 + g_2 \, g_3
~.
\eeq
$g_5$ is the 2-point function coefficient for the single chiral primary $\phi_5 = \phi_2 \, \phi_3$ 
at $\Delta = 5$. The explicit tree-level computation gives
\beq
\label{Cac}
g_3 = 5\, \NN_3^2 \frac{1}{({\rm Im}\tau)^3}
~,
\eeq
\beq
\label{Cad}
g_5 = \frac{35}{4}\, \NN_3^2 \frac{1}{({\rm Im}\tau)^5}
\eeq 
in agreement with the differential equation \eqref{Cab} for any $\NN_3$.

More involved examples with non-trivial degeneracy arise as we move up in scaling dimension.
For instance, at scaling dimension $\Delta=6$ there are two degenerate chiral primary fields
\beq
\label{Cae}
\phi_{(6,0)} = (\phi_2)^3
~, ~~
\phi_{(0,6)} = (\phi_3)^2
~.
\eeq
A tree-level computation shows that these fields are not orthogonal. 
The $2\times 2$ matrix of 2-point function coefficients is
\begin{align}
\label{Caf}
G_6 & = \begin{pmatrix}
		g_{(6,0)\overline{(6,0)}} & g_{(6,0)\overline{(0,6)}}\\
		g_{(0,6)\overline{(6,0)}} & g_{(0,6)\overline{(0,6)}}
	\end{pmatrix} 
	= \frac{1}{4 ({\rm{Im}}\,\tau)^6} \begin{pmatrix}
		45 & 15\, \NN_3^2 \\
		15 \, \NN_3^2 & 425 \, \NN_3^4 \\
	\end{pmatrix}
~.
\end{align}
Similarly, at scaling dimension $\Delta = 8$ there are two degenerate fields 
\beq
\label{Cag}
\phi_{(8,0)} = (\phi_2)^4~, ~~
\phi_{(2,6)} = \phi_2 \, (\phi_3)^2
\eeq
with the 2-point function coefficient matrix
\begin{align}
\label{Cai}
G_8 & = \begin{pmatrix}
		g_{(8,0)\overline{(8,0)}} & g_{(8,0)\overline{(2,6)}}\\
		g_{(2,6)\overline{(8,0)}} & g_{(2,6)\overline{(2,6)}}
	\end{pmatrix} = \frac{1}{4({\rm{Im}}\,\tau)^8} \begin{pmatrix}
		315 & 105 \, \NN_3^2 \\
		105\, \NN_3^2 & 1085\, \NN_3^4 
	\end{pmatrix}	
~.
\end{align}
The \tts equation \eqref{SU(N)af} at $\Delta =6$ is a matrix equation of the form
\beq
\label{Caj}
\frac{3}{2} \frac{1}{({\rm{Im}}\,\tau)^2} \delta^L_K = (G_8)_{(2,0) + K, \overline{(2,0)} + \bar R} (G_6)^{\bar R L} - 
(G_6)_{K \overline{(6,0)}} (g_4)^{-1} \delta^L_{(6,0)} - g_2\, \delta^L_K
~.
\eeq
One can verify that the algebraic equations \eqref{Caj} are satisfied by the tree-level 
expressions \eqref{Caf}, \eqref{Cai} and \eqref{Cda} for $N=3$, $n=1,2$.

\subsection*{$SU(N)$ observations}
\label{Ntree}

After the implementation of \eqref{treeSub} the $SU(N)$ \tts equations \eqref{SU(N)af} take the following 
algebraic form at tree-level
\beq
\label{Cba}
\frac{\Delta}{4} \tilde g^{\bar M_\Delta L_\Delta} \, \tilde g_{K_\Delta \, \bar M_\Delta}
= \tilde g_{K_\Delta +2, \bar R_\Delta + \bar 2} \, \tilde g^{\bar R_\Delta\, L_\Delta}
-\tilde g_{K_\Delta \, \bar R_\Delta} \, \tilde g^{\bar R_\Delta -\bar 2, L_\Delta -2}
-\tilde g_2\, \delta_{K_\Delta}^{L_\Delta}
~.
\eeq
This equation as an explicit index version of \eqref{RHStree} is the 
$SU(N)$ generalization of the $SU(3)$ matrix equation \eqref{Caj} above.
Although it is just a simple tree-level version of the full equations \eqref{treeSub} it continues to 
carry much of their complexity and encodes non-trivial information about the combinatorics of 
free field Wick contractions of 2-point functions of arbitrary multi-trace operators in the chiral ring.

Focusing on the 2-point functions $g_{2n}$ of the chiral primary fields $\phi_2 \propto (\Tr[\phi^2])^n$
we have observed experimentally (by direct {\sc Mathematica} computation of free field Wick contractions in a 
considerable range of values of $n, N$), that the following mathematical identity holds\footnote{We are not 
aware of a previous appearance of this identity in the literature. Related work that may be useful in proving it 
has appeared in \cite{Corley:2001zk}.}
\begin{align}
\label{Cda}
g_{2n} &= \frac{1}{16^n \left( {\rm Im}\tau \right)^{2n}}
\sum_{a_1,\ldots,a_{2n}=1}^{N^2-1} \delta_{a_1a_2} \cdots \delta_{a_{2n-1}a_{2n}}
\sum_{\sigma \in \SS_{2n}} \delta^{\sigma(a_1)\sigma(a_2)} \cdots \delta^{\sigma(a_{2n-1})\sigma(a_{2n})}
\nonumber\\
&= \frac{n!}{4^n} \left( \frac{N^2-1}{2} \right)_n \frac{1}{\left( {\rm Im}\tau \right)^{2n}}
~.
\end{align}
In this formula $(x)_n$ denotes the Pochhammer symbol
\beq
\label{Cdb}
(x)_n = x (x+1) \cdots (x+n-1)
~,
\eeq
$\SS_{2n}$ refers to the group of permutations of $2n$ elements and $\sigma$ is the generic permutation in this 
group. 

Although currently we do not have an analytic proof of this formula, we expect that it holds generally
for any value of the positive integers $n\geq 1$, $N>1$. For example, for $N=2$ (the $SU(2)$ case,
where there are no degeneracies and equations \eqref{2treeai} make up the full set of \tts equations)
one can easily see that the Pochhammer formula \eqref{Cda} reproduces the result \eqref{2treeae}, \eqref{2treeag}.
As another explicit check, notice that all the values of $g_{2n}$ (for $n=1,2,3,4$) in the previous $SU(3)$
section are consistent with \eqref{Cda}.
 
The intriguing fact about \eqref{Cda} is that it predicts values of $g_{2n}$ (at all $N>1$) that obey the tree-level 
version of the same semi-infinite Toda chain
\beq
\label{Cdc}
\partial_\tau \partial_{\bar \tau} \log g_{2n} = \frac{g_{2n+2}}{g_{2n}} - \frac{g_{2n}}{g_{2n-2}} -g_2
\eeq
that followed directly from the \tts equations in the $SU(2)$ case. This is not an obvious property of 
the matrix equations \eqref{Cba} at arbitrary $N$ and hints at a hidden underlying structure that will be useful 
to understand further. Moreover, if \eqref{Cdc} holds for $g_{2n}$ at all $N$ beyond tree-level it would allow us
to use the $SU(N)$ $S^4$ partition function to obtain a complete non-perturbative 
solution of the two-point functions $\left< \left(\phi_2 \right)^n (x) \left(\phi_2 \right)^n (0) \right>$ in the $SU(N)$
theory similar to the $SU(2)$ case above. These issues and their implications for the structure of 
the $SU(N)$ \tts equations (as well as possible extensions to more general chiral primary fields) are currently
under investigation.

\section{Summary and prospects}
\label{future}

We argued that the combination of supersymmetric localization techniques and exact relations like the \tts 
equations opens the interesting prospect for a new class of exact non-perturbative results in superconformal
field theories. 

In this paper we focused on four-dimensional $\NN=2$ superconformal
field theories. Combining the \tts equations of Ref.\ \cite{Papadodimas:2009eu} with the recent proposal
\cite{Gerchkovitz:2014gta} that relates the Zamolodchikov metric on the moduli space of $\NN=2$ SCFTs
to derivatives of the $S^4$ partition function we found useful exact relations between 2- and 3-point functions 
of $\NN=2$ chiral primary operators. In some cases, like the case of $SU(2)$ SCQCD, the \tts equations 
form a semi-infinite Toda chain and a unique solution can be determined easily in terms of the well-known
$S^4$ partition function of the $SU(2)$ theory. The solution provides exact formulae for the 2- and 3-point
functions of all the chiral primary fields of this theory as a function of the (complexified) gauge coupling.
We verified independently several aspects of this result with explicit computations
in perturbation theory up to 2-loops.

In more general situations, e.g. the $SU(N)$ SCQCD theory, the structure of the \tts equations is 
further complicated by the non-trivial mixing of degenerate chiral primary fields. We provided preliminary
observations of an underlying hidden structure in these equations that is worth investigating further.
The minimum data needed to determine a unique complete solution of the general $SU(N)$ \tts equations, 
and the structure of that solution, remains an interesting largely open question. 
It would be useful to know if a few fundamental general properties, like
positivity of 2-point functions over the entire conformal manifold, combined with some `boundary' data,
e.g. weak coupling perturbative data, are enough to specify a unique solution.

An exact solution of the \tts equations would have several important implications. In section \ref{extremal} we argued that the explicit knowledge of 2- and 3-point functions of chiral
primary operators can be used to determine also the generic extremal $n$-point correlation function
of these operators. In a different direction these results can also be used as input in a general bootstrap 
program in $\NN=2$ SCFTs to determine wider classes of correlation functions, spectral data $etc$.
Interesting work along similar lines appeared recently in \cite{Beem:2013sza}. For the case of $\NN=2$ SCQCD
we note that the methods developed in \cite{Beem:2013sza} (e.g. the correspondence with 
two-dimensional chiral algebras) are best suited for a discussion of the mesonic (Higgs branch) 
chiral primaries and are less useful for the $\NN=2$ (Coulomb branch) chiral primaries analyzed in 
the present paper. As a result, our approach can be viewed in this context as a different method providing useful 
complementary input.

In the main text we considered mostly the case of $\NN=2$ SCQCD theories as an illustrative example.
It would be interesting to extend the analysis to other four-dimensional $\NN=2$ theories, e.g. 
other Lagrangian theories, or the class $\SS$ theories \cite{Gaiotto:2009we,Alday:2009aq}.
Eventually, one would also like to move away from $\NN=2$ supersymmetry and explore situations
with less supersymmetry where quantum dynamics are known to exhibit a plethora of new effects. 
Two obvious hurdles in this direction are the following: $(i)$ it is known that the 
$S^4$ partition function of $\NN=1$ theories is ambiguous \cite{Gerchkovitz:2014gta}; $(ii)$ it is currently 
unknown whether there is any useful generalization of the \tts equations to $\NN=1$ theories 
\cite{Papadodimas:2009eu}. A related question has to do with the extension of these techniques to
theories of diverse amounts of supersymmetry in different dimensions,
e.g. three-dimensional SCFTs.

Originally, topological-antitopological fusion and the \tts equations \cite{Cecotti:1991me, Cecotti:1991vb} 
were also useful in analyzing two-dimensional $\NN=(2,2)$ massive theories. Therefore, another interesting 
direction is to explore whether a similar application of the \tts equations is also possible in four dimensions. Massive 
four-dimensional $\NN=2$ theories, like $\NN=2$ SYM theory, would be an interesting example. Related questions
were discussed in \cite{Cecotti:2013mba}.

\section*{Acknowledgments}

\noindent We would like to thank M. Buican, J. Drummond, M. Kelm,  W. Lerche, B. Pioline,
M. Rosso, D. Tong, C. Vergu, C. Vafa, C. Vollenweider and A. Zhedanov for useful discussions.
We used JaxoDraw \cite{Binosi:2003yf,Binosi:2008ig} to draw all the Feynman diagrams in this paper.
The work of M.B. is supported in part by a grant of the Swiss National Science Foundation.
The work of V.N. was supported in part by European Union's Seventh Framework
Programme under grant agreements (FP7-REGPOT-2012-2013-1) no 316165,
PIF-GA-2011-300984, the EU program ``Thales'' MIS 375734 and was also co-financed
by the European Union (European Social Fund, ESF) and Greek national funds through
the Operational Program ``Education and Lifelong Learning'' of the National Strategic
Reference Framework (NSRF) under ``Funding of proposals that have received a positive
evaluation in the 3rd and 4th Call of ERC Grant Schemes''. K.P. would like to thank the 
Royal Netherlands Academy of Sciences (KNAW).

\appendix
\addtocontents{toc}{\protect\setcounter{tocdepth}{1}}
\addtocontents{lof}{\protect\setcounter{tocdepth}{1}}

\section{Collection of useful facts about $S^4$ partition functions}
\label{SpecialFunctions}

In section \ref{sqcd} we make use of the $S^4$ partition function of $\NN=2$ $SU(N)$ SYM theories 
coupled to $2N$ hypermultiplets. Some of the pertinent details of this partition function are summarized
for the convenience of the reader in this appendix.

The $S^4$ partition function of $\NN=2$ gauge theories was computed using supersymmetric
localization in \cite{Pestun:2007rz} and the general result takes the form
\beq
\label{sfaa}
Z_{S^4}(\tau,\bar \tau) = \frac{1}{|\WW|}
\int da \, \Delta(a)\, Z_{\rm tree}(a) \, Z_{\rm 1-loop}(ia) \, \left| Z_{\rm inst}(ia,r^{-1},r^{-1},q) \right|^2~,
\eeq
where the integral is performed over the Cartan subalgebra of the gauge group ${\cal G}$, 
\begin{align}
\Delta(a)=\prod_{\alpha\, \in\,  {\rm roots}\, {\rm of}\, G} \alpha(a)
\end{align}
is the Vandermond determinant, 
$Z_{\rm tree}$ is the classical tree-level contribution, $Z_{\rm 1-loop}$ is the
1-loop contribution and $Z_{\rm inst}$ is Nekrasov's instanton partition function \cite{Nekrasov:2002qd}. 
$r$ denotes the radius of $S^4$ and $q=e^{2\pi i \tau}$. $|\WW|$ is the order of the Weyl group ${\cal G}$.

In the case of the $SU(N)$ $\NN=2$ SCQCD theories the elements of the 
Cartan subalgebra are parametrized by $N$ real parameters $a_i$ $(i=1,\ldots,N)$ 
satisfying the zero-trace condition $\sum_{i=1}^N a_i=0$, and
\beq
\label{sfab}
\Delta(a) = \prod_{i\neq j}^N (a_i - a_j)
~,
\eeq
\beq
\label{sfac}
Z_{\rm tree} = e^{-2 \pi \, {\rm Im}(\tau)\sum_{i=1}^{N} a_i^2}
~,
\eeq
\beq
\label{sfad}
Z_{\rm 1-loop}= \frac{\prod_{i\neq j} H(ir(a_i-a_j))}
{\prod_{j=1}^N \left( H(ira_j)H(-ira_j) \right)^{2N}}
~.\eeq
The instanton factor $Z_{\rm inst}$ has a more complicated form. General expressions can be found in 
\cite{Nekrasov:2002qd,Nekrasov:2003rj,Alday:2009aq}.
In the main text we set $r=1$ for the radius of $S^4$.

The special function $H$ that appears in the one-loop contribution is related
to the Barnes $G$-function \cite{Barnes}
\beq
\label{sfae}
G(1+z) = (2\pi)^{\frac{z}{2}} e^{-((1+\gamma z^2)+z)/2}
\prod_{n=1}^\infty \left( 1+\frac{z}{n} \right)^n 
e^{-z+\frac{z^2}{2n}}
\eeq
($\gamma$ is the Euler constant) through the defining equation
\beq
\label{sfag}
H(z) = G(1+z) G(1-z)
= e^{-(1+\gamma)z^2} \prod_{n=1}^\infty
\left( 1-\frac{z^2}{n^2} \right)^n \prod_{n=1}^\infty e^{\frac{z^2}{n}}
~.\eeq

\section{Conventions in $SU(N)$ $\NN=2$ SCQCD}
\label{conventions}

Here we collect our conventions for the $\NN=2$ SCQCD theories with gauge group $SU(N)$.

The $\NN=2$ chiral ring of the $SU(N)$ SCQCD theory is generated by the single-trace 
operators
\beq
\label{Aaa}
\phi_\ell \propto \Tr \big[ \varphi^\ell \big]~,~~ \ell = 2,3,\ldots, N
~.
\eeq
The descendant
\beq
\label{Aaaa}
\OO_\tau = \qq^4 \cdot \phi_2
\eeq
of the chiral primary $\phi_2$, that has the lowest scaling dimension $\Delta=2$, controls the 
exactly marginal deformation 
\beq
\label{Aab}
\delta S = \frac{\delta \tau}{4\pi^2} \int d^4x\, \OO_\tau(x) 
+ \frac{\delta \bar \tau}{4\pi^2} \int d^4 x \, \overline{\OO}_{\tau} (x)
~.
\eeq
The complex marginal coupling is $\tau = \frac{\theta}{2\pi}+\frac{4\pi i}{g_{YM}^2}$, and
we normalize the elementary fields of the theory so that the full Lagrangian in components takes the form
\beq
\label{Aaca}
\LL = \LL_{vector} + \LL_{hyper}
~,
\eeq
\begin{align}
\label{Aacb}
\LL_{vector} 
=& -\frac{1}{g_{YM}^2\, N} \Tr \bigg(
\frac{1}{4} F_{\mu\nu} F^{\mu\nu} +\frac{g_{YM}^2\, \theta}{32\pi^2} F_{\mu\nu} \tilde F^{\mu\nu} 
+ i \bar \lambda_\II \bar \sigma^\mu \DD_\mu \lambda^{\II} 
+ \DD_\mu \varphi\, \DD^\mu \bar \varphi
\nonumber\\
&+ i \sqrt 2 \left( \epsilon_{\II\JJ} \lambda^\II \lambda^\JJ \bar\varphi
 - \epsilon^{\II\JJ} \bar\lambda_\II \bar\lambda_\JJ \, \varphi  \right)
 +\frac{1}{2} \left[ \varphi,\bar \varphi \right]^2 \bigg)
 ~,
\end{align}
\begin{align}
\label{Aacc}
\LL_{hyper} =& - \bigg( \DD^\mu \bar Q^\II \DD_\mu Q_\II
+ i \bar\psi \bar\sigma^\mu \DD_\mu \psi + i \tilde \psi \bar\sigma^\mu \DD_\mu \bar{\tilde\psi}
\nonumber\\
&+i \sqrt 2 \Big( \epsilon^{\II\JJ} \bar\psi \bar\lambda_\II Q_\JJ - \epsilon_{\II\JJ} \bar Q^\II \lambda^\JJ \psi
+\tilde\psi \lambda^\II Q_\II - \bar Q^\II \bar\lambda_\II \bar{\tilde\psi}
+ \tilde\psi \varphi \psi - \bar\psi \bar\varphi \bar{\tilde\psi} \Big)
\nonumber\\
&+ \bar Q_\II \left(\bar\varphi \varphi +\varphi\bar\varphi \right) Q^\II
+g_{YM}^2 \VV(Q)
\bigg)
~,
\end{align}
with
\begin{align}
\label{Aacd}
\VV(Q) = &\left(\bar Q^{\II~i}_a Q_{\II~j}^{~a} \right)
\left(\bar Q^{\JJ~j}_b Q_{\JJ~i}^{~b} \right)
-\frac{1}{2} \left(\bar Q^{\II~i}_a Q_{\JJ~j}^{~a} \right)
\left(\bar Q^{\JJ~j}_b Q_{\II~i}^{~b} \right)
\nonumber\\
&+\frac{1}{N} \left( \frac{1}{2} \left(\bar Q^{\II~i}_a Q_{\II~i}^{~a} \right)
\left(\bar Q^{\JJ~j}_b Q_{\JJ~j}^{~b} \right)
- \left(\bar Q^{\II~i}_a Q_{\JJ~i}^{~a} \right)
\left(\bar Q^{\JJ~j}_b Q_{\II~j}^{~b} \right)
\right)
\end{align}
the D-term potential for the hypermultiplet complex scalars $Q$. 

We use standard notation where
\beq
\label{Aad}
F_{\mu\nu} = F_{\mu\nu}^a T^a~, ~~
F_{\mu\nu}^a = \partial_\mu A_\nu^a -\partial_\nu A_\mu^a + f^{abc} A_\mu^b A_\nu^c~, ~~
\tilde F_{\mu\nu}^a = \frac{1}{2} \varepsilon_{\mu\nu\rho\sigma} F^{\rho\sigma}
\eeq
and $T^a$ $(a=1,\ldots,N^2-1)$ is a basis of the Lie algebra generators of $SU(N)$
with the normalization 
\beq
\label{Aadb}
\Tr \left[ T_{adj}^a T_{adj}^b \right] = N\, \delta^{ab}
~,~~
\Tr \left[ T_{\Box}^a T_{\Box}^b \right] =\frac{1}{2} \delta^{ab}
\eeq
in the adjoint and fundamental representations respectively.
The gauge-covariant derivatives are
\beq
\label{Aada}
\DD_\mu =\partial_\mu + i A_\mu 
~.
\eeq
$\II,\JJ,\ldots = \pm$ are $SU(2)_R$ indices raised and lowered with the antisymmetric symbols 
$\epsilon_{\II\JJ}$. $i,j=1,\ldots, 2N$ in \eqref{Aacd} are flavor indices. 
The $\NN=2$ vector fields in the adjoint representation include the
bosons $A_\mu, \varphi$ and the fermions $\lambda^\II$. The $2N$ $\NN=2$ hypermultiplet fields
in the fundamental representation include $2N$ complex bosons $Q_\II$ and $2N$ fermion doublets 
$(\psi, \tilde \psi)$.

In this normalization all the $\tau$ dependence is loaded on the vector part of the Lagrangian.\footnote{The
last term of the hypermultiplet interactions, $g_{YM}^2 \VV(Q)$, appears to be $g_{YM}$-dependent, but
this is only so after we integrate out the $D$ auxiliary field. Before integrating out
$D$ the Lagrangian $\LL_{vector}$ has a term $\frac{1}{2g_{YM}^2}D^2$ and $\LL_{hyper}$ has no
explicit $g_{YM}$-dependence.} This is consistent with \eqref{Aaaa}, \eqref{Aab} and the identification 
\begin{align}
\label{Aae}
\OO_\tau = \frac{i\pi}{2N} \Tr \bigg[ &\frac{1}{8} F_{\mu\nu+} F^{\mu\nu+} 
+ i \bar \lambda_{\II} \bar\sigma^\mu \DD_\mu \lambda^{\II}
-\bar\varphi \, \DD_\mu \DD^\mu \varphi 
- \frac{1}{2} D^2 - \bar F  F
\nonumber\\
& +\sqrt 2 \left( \epsilon_{\II\JJ} \lambda^{\II} \lambda^{\JJ} \bar\varphi
-\epsilon^{\II\JJ} \bar\lambda_{\II} \bar\lambda_{\JJ} \varphi \right) 
- D[\varphi,\bar\varphi] \bigg]
~.
\end{align}
$F_{\mu\nu\pm}=F_{\mu\nu} \mp i \tilde F_{\mu\nu}$ is the (anti)self-dual part of the
gauge field strength. $D$ and $F$ are respectively the $D$- and $F$-auxiliary fields
of the $\NN=1$ vector and $\NN=1$ chiral multiplet that make up the $\NN=2$ vector multiplet.

\section{An (eccentric) proof of equation \eqref{grouplemma}}
\label{LemmaApp}

In this section we will give a proof of equation \eqref{grouplemma}. Instead of giving a direct combinatoric proof, 
we will proceed as follows. Consider the ${\cal N}=4$ SYM theory with gauge group ${\cal G}$, in the free limit. This 
can also be thought of as an ${\cal N}=2$ SCFT. This theory has 6 real scalars $\Phi^{\rm I}, {\rm I} =1,...,6.$ We 
consider the complex combination
\be
\varphi = \Phi^1 + i \Phi^2 .
\ee
The chiral primary, whose descendant is the marginal operator, has the form
\be
\phi_2 = {\cal N}\, \Tr[\varphi^2] .
\ee
where the normalization constant ${\cal N}$ was determined in previous sections.
We define the 2-point function 
\be
g_{2\overline{2}} = \langle \phi_2 \overline{\phi}_2\rangle.
\ee
A general chiral primary of charge $R$ can be written as a multitrace operator of the form
\be
\phi_K \propto \Tr[\varphi^{n_1}]...\Tr[\varphi^{n_k}] ,
\ee
where $2\sum n_i = R$. The trace is taken in the adjoint of ${\cal G}$. Similarly we define the anti-chiral 
primaries and the matrix of 2-point functions $g_{K\overline{L}} = \langle \phi_K \overline{\phi}_L\rangle$. Notice 
that the matrix of 2-point functions $g_{K\overline{L}}$ is not diagonal in the basis of multitrace operators and is 
somewhat cumbersome to compute by considering Wick contractions.

Our starting point is to consider the following 4-point function 
\be
\label{4pointid}
{\rm A} = \langle  \phi_2(x_1)\, \overline{\phi}_2 (x_2)\,\phi_K(x_3)\, \overline{\phi}_L(x_4)\rangle .
\ee
Here $K,L$ can be different chiral primaries, but by R-charge conservation this 4-point function is nonzero only if 
$K,L$ have the same $R$-charge. By Wick contractions it is not hard to see that there are only three possible 
structures of the coordinate dependence for this correlator. So the general form is
\be
\label{4pointbb}
{\rm A} = {p_1 \over |x_{12}|^4 |x_{34}|^{2\Delta_K}} 
+ {p_2 \over |x_{12}|^2 |x_{14}|^2 |x_{23}|^2|x_{34}|^{2\Delta_K-2}} 
+ {p_3 \over |x_{14}|^4 |x_{23}|^4 |x_{34}|^{2\Delta_K-4}} .
\ee
In principle we can compute the constants $p_1,p_2,p_3$ by working out the combinatorics of the Wick 
contractions, however we will try to avoid this. By considering the double OPE in the $(13)\rightarrow (24)$ 
channel we learn that
\be
\label{OPE1}
p_1+p_2+p_3 = C_{2K}^P g_{P\overline{Q}} C^{*\overline{Q}}_{\overline{2} \overline{L}} .
\ee
By considering the OPE in the $(12)\rightarrow (34)$ channel we have 
\be
\label{OPE2}
p_1 = g_{2\overline{2}}\,\, g_{K\overline{L}} .
\ee
Finally from the OPE in the $(14)\rightarrow (23)$ channel we find
\be
\label{OPE3}
p_3 = g_{K\overline{N}} C^{* \overline{N}}_{\overline{2} \overline{U}} g^{\overline{U}V} C_{2V}^R g_{R 
\overline{L}} .
\ee
Using these results we have completely fixed the 4-point function \eqref{4pointid} in the free limit, in terms of the 
2- and 3-point function coefficients which enter the \tts equations.

However, the desired equation \eqref{grouplemma} expresses a nontrivial relation among these coefficients. We 
will now argue that the consistency of the underlying CFT implies the desired relation.

We will establish the relation by the following argument. The tree level correlator \eqref{4pointid} can be thought of 
as a correlator in a theory of only ${\rm dim}\,{\cal G}$ complex scalar fields\footnote{Since we are in the free limit 
the presence of the other fields does not make any difference to the counting of the Wick combinatorics.}. This by 
itself is a consistent conformal field theory with a central charge $c_{\rm scalar}$ which is related to 
${\rm dim}\,{\cal G}$ by
\be
c_{\rm scalar} = {8 \over 3} {\rm dim} {\cal G} .
\ee
To derive equation \eqref{OPE2} we considered the OPE in the channel $(12)\rightarrow (34)$ and only kept the 
leading term, i.e. the identity operator. One of the subleading contributions involves conformal block of the stress 
energy tensor. In any consistent CFT the contribution of this block is fully determined using Ward identities, by the 
central charge of the CFT and by the conformal dimension of the external operators \cite{Dolan:2002zh}. Our 
strategy is to:
\begin{itemize}
\item[$a)$] isolate the contribution of the conformal block of the stress energy
tensor for the 4-point function \eqref{4pointid}, \eqref{4pointbb} written in terms of the data \eqref{OPE1}, 
\eqref{OPE2}, \eqref{OPE3} and 
\item[$b)$] demand that this contribution
is the same as that predicted by general arguments based on the Ward identities for  CFTs. We will discover that 
this requirement leads to the desired formula \eqref{grouplemma}.
\end{itemize}

We write equation \eqref{4pointbb} in a notation which is somewhat more convenient to perform the conformal 
block expansion
\be
\label{docpw}
{\rm A} = 
 {1\over |x_{12}|^4 |x_{34}|^{2\Delta_K}} \left( p_1 + p_2 {u \over v} + p_3  {u^2 \over v^2} \right) ,
\ee
where we have introduced the conformal cross ratios
\be
u = {|x_{12}|^2 |x_{34}|^2 \over |x_{13}|^2 |x_{24}|^2}\qquad,\qquad v
= {|x_{14}|^2 |x_{23}|^2 \over |x_{13}|^2 |x_{24}|^2} .
\ee
It is easy to see that the term $p_1 = g_{2\overline{2}} g_{K \overline{L}}$ is coming from the exchange of the 
identity operator (the reason that it is not equal to 1 is because our 2-point functions are not normalized to be $\propto{1\over |x|^{2\Delta}}$). 
With a little work on the conformal block expansion in the $u\rightarrow 0 , v\rightarrow 1$ channel, we find that the 
block of the stress tensor comes with the coefficient
\be
\label{docpw}
{\rm A} = 
 {1\over |x_{12}|^4 |x_{34}|^{2\Delta_K}} \left( \ldots + {2\over 3}p_2 u G^{(2)}(1,1,4,u,v)  + \ldots\right) ,
\ee
where the function $G^{(2)}$ is defined in \cite{Dolan:2002zh}.

On the other hand, the Ward identities predict \cite{Dolan:2002zh} that for any consistent CFT if we have the 4-point 
function $\langle \phi (x_1) \phi (x_2) \phi' (x_3) \phi'(x_4)\rangle$ and we expand it in the $(12)\rightarrow (34)$ 
channel, then the stress tensor must contribute like
\be
\langle \phi (x_1) \phi (x_2) \phi' (x_3) \phi'(x_4)\rangle = {1\over |x_{12}|^4 |x_{34}|^{2\Delta_K}} 
\left( \ldots + {16 \Delta \Delta' \over 9 c} g_{\phi\phi} g_{\phi'\phi'}  u G^{(2)}(1,1,4,u,v) +  \ldots\right) ,
\ee
where $c$ is the central charge of the CFT and $g_{\phi\phi}, g_{\phi'\phi'}$ is the normalization of the 2-point 
functions, which in \cite{Dolan:2002zh} was taken to be 1. In our case we have
$\Delta =2, \Delta' = \Delta_K = R/2$ and $c \equiv c_{\rm scalar} ={8\over 3} {\rm dim} {\cal G}$. 
Putting everything together we find that what we {\it expect} in a consistent CFT for the 4-point function 
\eqref{4pointbb} is that the stress-tensor conformal block comes with the coefficient
\be
\label{docpw2}
{\rm A} = 
{1\over |x_{12}|^4 |x_{34}|^{2\Delta_K}} 
\left( \ldots + {2\over 3}{ R \over {\rm dim}\,{\cal G}}\,p_1\, u G^{(2)}(1,1,4,u,v)  + \ldots\right) .
\ee
Comparing this to what we found in \eqref{docpw} we conclude that consistency of the CFT demands the relation
\be
p_2 = { R\over {\rm dim}\, {\cal G}}\, p_1 .
\ee
Using the expression \eqref{OPE1}, \eqref{OPE2} and \eqref{OPE3} it is straightforward to show that this implies
\be
\label{finallemma}
-[C_2,\overline{C}_{\overline{2}}]_K^L + g_{2 \overline{2}} \delta_K^L \left(1 + {R \over {\rm dim}\,{\cal G}}  \right) =0 .
\ee

\bibliographystyle{utphys}
\bibliography{ttstarpaper}

\providecommand{\href}[2]{#2}\begingroup\raggedright\begin{thebibliography}{10}

\bibitem{Leigh:1995ep}
R.~G. Leigh and M.~J. Strassler, ``{Exactly marginal operators and duality in
  four-dimensional N=1 supersymmetric gauge theory},''
  \href{http://dx.doi.org/10.1016/0550-3213(95)00261-P}{{\em Nucl.Phys.}
  {\bfseries B447} (1995) 95--136},
\href{http://arxiv.org/abs/hep-th/9503121}{{\ttfamily arXiv:hep-th/9503121
  [hep-th]}}.

\bibitem{Cecotti:1991me}
S.~Cecotti and C.~Vafa, ``{Topological antitopological fusion},''
\href{http://dx.doi.org/10.1016/0550-3213(91)90021-O}{{\em Nucl.Phys.}
  {\bfseries B367} (1991) 359--461}.

\bibitem{Cecotti:1991vb}
S.~Cecotti and C.~Vafa, ``{Exact results for supersymmetric sigma models},''
  \href{http://dx.doi.org/10.1103/PhysRevLett.68.903}{{\em Phys.Rev.Lett.}
  {\bfseries 68} (1992) 903--906},
\href{http://arxiv.org/abs/hep-th/9111016}{{\ttfamily arXiv:hep-th/9111016
  [hep-th]}}.

\bibitem{Papadodimas:2009eu}
K.~Papadodimas, ``{Topological Anti-Topological Fusion in Four-Dimensional
  Superconformal Field Theories},''
  \href{http://dx.doi.org/10.1007/JHEP08(2010)118}{{\em JHEP} {\bfseries 1008}
  (2010) 118},
\href{http://arxiv.org/abs/0910.4963}{{\ttfamily arXiv:0910.4963 [hep-th]}}.

\bibitem{Cecotti:2013mba}
S.~Cecotti, D.~Gaiotto, and C.~Vafa, ``{$tt^*$ geometry in 3 and 4
  dimensions},'' \href{http://dx.doi.org/10.1007/JHEP05(2014)055}{{\em JHEP}
  {\bfseries 1405} (2014) 055},
\href{http://arxiv.org/abs/1312.1008}{{\ttfamily arXiv:1312.1008 [hep-th]}}.

\bibitem{Vafa:2014lca}
C.~Vafa, ``{tt* Geometry and a Twistorial Extension of Topological Strings},''
\href{http://arxiv.org/abs/1402.2674}{{\ttfamily arXiv:1402.2674 [hep-th]}}.

\bibitem{Gerchkovitz:2014gta}
E.~Gerchkovitz, J.~Gomis, and Z.~Komargodski, ``{Sphere Partition Functions and
  the Zamolodchikov Metric},''
\href{http://arxiv.org/abs/1405.7271}{{\ttfamily arXiv:1405.7271 [hep-th]}}.

\bibitem{Pestun:2007rz}
V.~Pestun, ``{Localization of gauge theory on a four-sphere and supersymmetric
  Wilson loops},'' \href{http://dx.doi.org/10.1007/s00220-012-1485-0}{{\em
  Commun.Math.Phys.} {\bfseries 313} (2012) 71--129},
\href{http://arxiv.org/abs/0712.2824}{{\ttfamily arXiv:0712.2824 [hep-th]}}.

\bibitem{Lee:1998bxa}
S.~Lee, S.~Minwalla, M.~Rangamani, and N.~Seiberg, ``{Three point functions of
  chiral operators in D = 4, N=4 SYM at large N},'' {\em Adv.Theor.Math.Phys.}
  {\bfseries 2} (1998) 697--718,
\href{http://arxiv.org/abs/hep-th/9806074}{{\ttfamily arXiv:hep-th/9806074
  [hep-th]}}.

\bibitem{D'Hoker:1998tz}
E.~D'Hoker, D.~Z. Freedman, and W.~Skiba, ``{Field theory tests for correlators
  in the AdS / CFT correspondence},''
  \href{http://dx.doi.org/10.1103/PhysRevD.59.045008}{{\em Phys.Rev.}
  {\bfseries D59} (1999) 045008},
\href{http://arxiv.org/abs/hep-th/9807098}{{\ttfamily arXiv:hep-th/9807098
  [hep-th]}}.

\bibitem{D'Hoker:1999ea}
E.~D'Hoker, D.~Z. Freedman, S.~D. Mathur, A.~Matusis, and L.~Rastelli,
  ``{Extremal correlators in the AdS / CFT correspondence},''
\href{http://arxiv.org/abs/hep-th/9908160}{{\ttfamily arXiv:hep-th/9908160
  [hep-th]}}.

\bibitem{Intriligator:1998ig}
K.~A. Intriligator, ``{Bonus symmetries of N=4 superYang-Mills correlation
  functions via AdS duality},''
  \href{http://dx.doi.org/10.1016/S0550-3213(99)00242-4}{{\em Nucl.Phys.}
  {\bfseries B551} (1999) 575--600},
\href{http://arxiv.org/abs/hep-th/9811047}{{\ttfamily arXiv:hep-th/9811047
  [hep-th]}}.

\bibitem{Intriligator:1999ff}
K.~A. Intriligator and W.~Skiba, ``{Bonus symmetry and the operator product
  expansion of N=4 SuperYang-Mills},''
  \href{http://dx.doi.org/10.1016/S0550-3213(99)00430-7}{{\em Nucl.Phys.}
  {\bfseries B559} (1999) 165--183},
\href{http://arxiv.org/abs/hep-th/9905020}{{\ttfamily arXiv:hep-th/9905020
  [hep-th]}}.

\bibitem{Eden:1999gh}
B.~Eden, P.~S. Howe, and P.~C. West, ``{Nilpotent invariants in N=4 SYM},''
  \href{http://dx.doi.org/10.1016/S0370-2693(99)00705-4}{{\em Phys.Lett.}
  {\bfseries B463} (1999) 19--26},
\href{http://arxiv.org/abs/hep-th/9905085}{{\ttfamily arXiv:hep-th/9905085
  [hep-th]}}.

\bibitem{Petkou:1999fv}
A.~Petkou and K.~Skenderis, ``{A Nonrenormalization theorem for conformal
  anomalies},'' \href{http://dx.doi.org/10.1016/S0550-3213(99)00514-3}{{\em
  Nucl.Phys.} {\bfseries B561} (1999) 100--116},
\href{http://arxiv.org/abs/hep-th/9906030}{{\ttfamily arXiv:hep-th/9906030
  [hep-th]}}.

\bibitem{Howe:1999hz}
P.~S. Howe, C.~Schubert, E.~Sokatchev, and P.~C. West, ``{Explicit construction
  of nilpotent covariants in N=4 SYM},''
  \href{http://dx.doi.org/10.1016/S0550-3213(99)00768-3}{{\em Nucl.Phys.}
  {\bfseries B571} (2000) 71--90},
\href{http://arxiv.org/abs/hep-th/9910011}{{\ttfamily arXiv:hep-th/9910011
  [hep-th]}}.

\bibitem{Heslop:2001gp}
P.~Heslop and P.~S. Howe, ``{OPEs and three-point correlators of protected
  operators in N=4 SYM},''
  \href{http://dx.doi.org/10.1016/S0550-3213(02)00023-8}{{\em Nucl.Phys.}
  {\bfseries B626} (2002) 265--286},
\href{http://arxiv.org/abs/hep-th/0107212}{{\ttfamily arXiv:hep-th/0107212
  [hep-th]}}.

\bibitem{Baggio:2012rr}
M.~Baggio, J.~de~Boer, and K.~Papadodimas, ``{A non-renormalization theorem for
  chiral primary 3-point functions},''
  \href{http://dx.doi.org/10.1007/JHEP07(2012)137}{{\em JHEP} {\bfseries 1207}
  (2012) 137},
\href{http://arxiv.org/abs/1203.1036}{{\ttfamily arXiv:1203.1036 [hep-th]}}.

\bibitem{shortpaper}
M.~Baggio, V.~Niarchos, and K.~Papadodimas, ``{Exact correlation functions in
  $SU(2)$ ${\cal N}=2$ superconformal QCD},''.

\bibitem{Dolan:2002zh}
F.~Dolan and H.~Osborn, ``{On short and semi-short representations for
  four-dimensional superconformal symmetry},''
  \href{http://dx.doi.org/10.1016/S0003-4916(03)00074-5}{{\em Annals Phys.}
  {\bfseries 307} (2003) 41--89},
\href{http://arxiv.org/abs/hep-th/0209056}{{\ttfamily arXiv:hep-th/0209056
  [hep-th]}}.

\bibitem{Buican:2014qla}
M.~Buican, T.~Nishinaka, and C.~Papageorgakis, ``{Constraints on Chiral
  Operators in N=2 SCFTs},''
\href{http://arxiv.org/abs/1407.2835}{{\ttfamily arXiv:1407.2835 [hep-th]}}.

\bibitem{Lerche:1989uy}
W.~Lerche, C.~Vafa, and N.~P. Warner, ``{Chiral Rings in N=2 Superconformal
  Theories},''
\href{http://dx.doi.org/10.1016/0550-3213(89)90474-4}{{\em Nucl.Phys.}
  {\bfseries B324} (1989) 427}.

\bibitem{Seiberg:1988pf}
N.~Seiberg, ``{Observations on the Moduli Space of Superconformal Field
  Theories},''
\href{http://dx.doi.org/10.1016/0550-3213(88)90183-6}{{\em Nucl.Phys.}
  {\bfseries B303} (1988) 286}.

\bibitem{Kutasov:1988xb}
D.~Kutasov, ``{Geometry on the Space of Conformal Field Theories and Contact
  Terms},''
\href{http://dx.doi.org/10.1016/0370-2693(89)90028-2}{{\em Phys.Lett.}
  {\bfseries B220} (1989) 153}.

\bibitem{Strominger:1990pd}
A.~Strominger, ``{SPECIAL GEOMETRY},''
\href{http://dx.doi.org/10.1007/BF02096559}{{\em Commun.Math.Phys.} {\bfseries
  133} (1990) 163--180}.

\bibitem{Bershadsky:1993cx}
M.~Bershadsky, S.~Cecotti, H.~Ooguri, and C.~Vafa, ``{Kodaira-Spencer theory of
  gravity and exact results for quantum string amplitudes},''
  \href{http://dx.doi.org/10.1007/BF02099774}{{\em Commun.Math.Phys.}
  {\bfseries 165} (1994) 311--428},
\href{http://arxiv.org/abs/hep-th/9309140}{{\ttfamily arXiv:hep-th/9309140
  [hep-th]}}.

\bibitem{Ranganathan:1993vj}
K.~Ranganathan, H.~Sonoda, and B.~Zwiebach, ``{Connections on the state space
  over conformal field theories},''
  \href{http://dx.doi.org/10.1016/0550-3213(94)90436-7}{{\em Nucl.Phys.}
  {\bfseries B414} (1994) 405--460},
\href{http://arxiv.org/abs/hep-th/9304053}{{\ttfamily arXiv:hep-th/9304053
  [hep-th]}}.

\bibitem{deBoer:2008ss}
J.~de~Boer, J.~Manschot, K.~Papadodimas, and E.~Verlinde, ``{The Chiral ring of
  AdS(3)/CFT(2) and the attractor mechanism},''
  \href{http://dx.doi.org/10.1088/1126-6708/2009/03/030}{{\em JHEP} {\bfseries
  0903} (2009) 030},
\href{http://arxiv.org/abs/0809.0507}{{\ttfamily arXiv:0809.0507 [hep-th]}}.

\bibitem{Witten:1989ig}
E.~Witten, ``{On the Structure of the Topological Phase of Two-dimensional
  Gravity},''
\href{http://dx.doi.org/10.1016/0550-3213(90)90449-N}{{\em Nucl.Phys.}
  {\bfseries B340} (1990) 281--332}.

\bibitem{Dijkgraaf:1990dj}
R.~Dijkgraaf, H.~L. Verlinde, and E.~P. Verlinde, ``{Topological strings in d
  $<$ 1},''
\href{http://dx.doi.org/10.1016/0550-3213(91)90129-L}{{\em Nucl.Phys.}
  {\bfseries B352} (1991) 59--86}.

\bibitem{Dijkgraaf:1990qw}
R.~Dijkgraaf, H.~L. Verlinde, and E.~P. Verlinde,
``{Notes on topological string theory and 2-D quantum gravity},''.

\bibitem{Gaiotto:2009we}
D.~Gaiotto, ``{N=2 dualities},''
  \href{http://dx.doi.org/10.1007/JHEP08(2012)034}{{\em JHEP} {\bfseries 1208}
  (2012) 034},
\href{http://arxiv.org/abs/0904.2715}{{\ttfamily arXiv:0904.2715 [hep-th]}}.

\bibitem{Alday:2009aq}
L.~F. Alday, D.~Gaiotto, and Y.~Tachikawa, ``{Liouville Correlation Functions
  from Four-dimensional Gauge Theories},''
  \href{http://dx.doi.org/10.1007/s11005-010-0369-5}{{\em Lett.Math.Phys.}
  {\bfseries 91} (2010) 167--197},
\href{http://arxiv.org/abs/0906.3219}{{\ttfamily arXiv:0906.3219 [hep-th]}}.

\bibitem{Gadde:2009dj}
A.~Gadde, E.~Pomoni, and L.~Rastelli, ``{The Veneziano Limit of N = 2
  Superconformal QCD: Towards the String Dual of N = 2 SU(N(c)) SYM with N(f) =
  2 N(c)},''
\href{http://arxiv.org/abs/0912.4918}{{\ttfamily arXiv:0912.4918 [hep-th]}}.

\bibitem{Barnes}
E.~W. Barnes, ``{The theory of the double gamma function},'' {\em Philosophical
  Transactions of the Royal Society of London. Series A, Containing Papers of a
  Mathematical or Physical Character} {\bfseries 196} (1901) 265--387.

\bibitem{Nekrasov:2002qd}
N.~A. Nekrasov, ``{Seiberg-Witten prepotential from instanton counting},''
  \href{http://dx.doi.org/10.4310/ATMP.2003.v7.n5.a4}{{\em
  Adv.Theor.Math.Phys.} {\bfseries 7} (2004) 831--864},
\href{http://arxiv.org/abs/hep-th/0206161}{{\ttfamily arXiv:hep-th/0206161
  [hep-th]}}.

\bibitem{Russo:2012kj}
J.~G. Russo, ``{A Note on perturbation series in supersymmetric gauge
  theories},'' \href{http://dx.doi.org/10.1007/JHEP06(2012)038}{{\em JHEP}
  {\bfseries 1206} (2012) 038},
\href{http://arxiv.org/abs/1203.5061}{{\ttfamily arXiv:1203.5061 [hep-th]}}.

\bibitem{Gomis:2009xg}
J.~Gomis and T.~Okuda, ``{S-duality, 't Hooft operators and the operator
  product expansion},''
  \href{http://dx.doi.org/10.1088/1126-6708/2009/09/072}{{\em JHEP} {\bfseries
  0909} (2009) 072},
\href{http://arxiv.org/abs/0906.3011}{{\ttfamily arXiv:0906.3011 [hep-th]}}.

\bibitem{Andree:2010na}
R.~Andree and D.~Young, ``{Wilson Loops in N=2 Superconformal Yang-Mills
  Theory},'' \href{http://dx.doi.org/10.1007/JHEP09(2010)095}{{\em JHEP}
  {\bfseries 1009} (2010) 095},
\href{http://arxiv.org/abs/1007.4923}{{\ttfamily arXiv:1007.4923 [hep-th]}}.

\bibitem{Penati:1999ba}
S.~Penati, A.~Santambrogio, and D.~Zanon, ``{Two point functions of chiral
  operators in N=4 SYM at order g**4},''
  \href{http://dx.doi.org/10.1088/1126-6708/1999/12/006}{{\em JHEP} {\bfseries
  9912} (1999) 006},
\href{http://arxiv.org/abs/hep-th/9910197}{{\ttfamily arXiv:hep-th/9910197
  [hep-th]}}.

\bibitem{Penati:2000zv}
S.~Penati, A.~Santambrogio, and D.~Zanon, ``{More on correlators and contact
  terms in N=4 SYM at order g**4},''
  \href{http://dx.doi.org/10.1016/S0550-3213(00)00633-7}{{\em Nucl.Phys.}
  {\bfseries B593} (2001) 651--670},
\href{http://arxiv.org/abs/hep-th/0005223}{{\ttfamily arXiv:hep-th/0005223
  [hep-th]}}.

\bibitem{Chetyrkin:1981qh}
K.~Chetyrkin and F.~Tkachov, ``{Integration by Parts: The Algorithm to
  Calculate beta Functions in 4 Loops},''
\href{http://dx.doi.org/10.1016/0550-3213(81)90199-1}{{\em Nucl.Phys.}
  {\bfseries B192} (1981) 159--204}.

\bibitem{Gehrmann:2010ue}
T.~Gehrmann, E.~Glover, T.~Huber, N.~Ikizlerli, and C.~Studerus, ``{Calculation
  of the quark and gluon form factors to three loops in QCD},''
  \href{http://dx.doi.org/10.1007/JHEP06(2010)094}{{\em JHEP} {\bfseries 1006}
  (2010) 094},
\href{http://arxiv.org/abs/1004.3653}{{\ttfamily arXiv:1004.3653 [hep-ph]}}.

\bibitem{Smirnov:2008iw}
A.~Smirnov, ``{Algorithm FIRE -- Feynman Integral REduction},''
  \href{http://dx.doi.org/10.1088/1126-6708/2008/10/107}{{\em JHEP} {\bfseries
  0810} (2008) 107},
\href{http://arxiv.org/abs/0807.3243}{{\ttfamily arXiv:0807.3243 [hep-ph]}}.

\bibitem{Corley:2001zk}
S.~Corley, A.~Jevicki, and S.~Ramgoolam, ``{Exact correlators of giant
  gravitons from dual N=4 SYM theory},'' {\em Adv.Theor.Math.Phys.} {\bfseries
  5} (2002) 809--839,
\href{http://arxiv.org/abs/hep-th/0111222}{{\ttfamily arXiv:hep-th/0111222
  [hep-th]}}.

\bibitem{Beem:2013sza}
C.~Beem, M.~Lemos, P.~Liendo, W.~Peelaers, L.~Rastelli, {\em et~al.},
  ``{Infinite Chiral Symmetry in Four Dimensions},''
\href{http://arxiv.org/abs/1312.5344}{{\ttfamily arXiv:1312.5344 [hep-th]}}.

\bibitem{Binosi:2003yf}
D.~Binosi and L.~Theussl, ``{JaxoDraw: A Graphical user interface for drawing
  Feynman diagrams},'' \href{http://dx.doi.org/10.1016/j.cpc.2004.05.001}{{\em
  Comput.Phys.Commun.} {\bfseries 161} (2004) 76--86},
\href{http://arxiv.org/abs/hep-ph/0309015}{{\ttfamily arXiv:hep-ph/0309015
  [hep-ph]}}.

\bibitem{Binosi:2008ig}
D.~Binosi, J.~Collins, C.~Kaufhold, and L.~Theussl, ``{JaxoDraw: A Graphical
  user interface for drawing Feynman diagrams. Version 2.0 release notes},''
  \href{http://dx.doi.org/10.1016/j.cpc.2009.02.020}{{\em Comput.Phys.Commun.}
  {\bfseries 180} (2009) 1709--1715},
\href{http://arxiv.org/abs/0811.4113}{{\ttfamily arXiv:0811.4113 [hep-ph]}}.

\bibitem{Nekrasov:2003rj}
N.~Nekrasov and A.~Okounkov, ``{Seiberg-Witten theory and random partitions},''
\href{http://arxiv.org/abs/hep-th/0306238}{{\ttfamily arXiv:hep-th/0306238
  [hep-th]}}.

\end{thebibliography}\endgroup
\end{document}